\documentclass[reprint,prx
,longbibliography,nofootinbib]{revtex4-2}
\usepackage{amssymb}
\usepackage{amsmath}
\usepackage{braket}
\usepackage{color}
\usepackage[amsmath]{empheq}
\usepackage{tikz}
\usepackage{color}
\usepackage{nicefrac}
\usepackage{bbm}
\usepackage{tensor}
\usepackage{tikzsymbols}
\usepackage{ifsym}
\usepackage{graphicx}
\usepackage{tikz-cd}
\usepackage[pdftex, hidelinks]{hyperref}
\usepackage{natbib}
\usepackage{cancel}

\usepackage[shortlabels]{enumitem}
\setlist[description]{font=\normalfont}

\usepackage{graphicx}

\newcommand{\be}{\begin{equation}}
\newcommand{\ee}{\end{equation}}
\newcommand{\ba}{\begin{aligned}}
\newcommand{\ea}{\end{aligned}}

\begin{document}
\title{
Quantum jamming brings quantum mechanics to macroscopic scales
}
\author{Maurizio Fagotti}
\affiliation{Universit\'e Paris-Saclay, CNRS, LPTMS, 91405, Orsay, France}
\begin{abstract}
A quantum spin-$\frac{1}{2}$ chain with an axial symmetry is normally described by quasiparticles associated with the spins oriented along the axis of rotation. Kinetic constraints can enrich such a description by setting apart different species of quasiparticles, which can get stuck at high enough density, realising the quantum analogue of jamming. We identify a family of interactions satisfying simple kinetic constraints and consider generic translationally invariant models built up from them. 
We study dynamics following  a local unjamming perturbation in a jammed state.  
We show that they can be mapped into dynamics of ordinary unconstrained systems, 
but the nonlocality of the mapping changes the scales at which the phenomena manifest themselves. Scattering of quasiparticles, formation of bound states,  eigenstate localisation become all visible at macroscopic scales.  Depending on whether a symmetry is present or not,  the microscopic details of the jammed state turn out to have either a marginal or a strong effect. In the former case or when the initial state is almost homogeneous, we show that even a product state is turned into a macroscopic quantum~state.
\end{abstract}
\maketitle
\section{Introduction}
Low dimensional quantum many-body systems easily capture the interest of theoretical physicists as a result of their unusual properties, e.g., strong correlations~\cite{Giamarchi2003Quantum}, 
integrability in the presence of interactions~\cite{Essler2005The} in 1D, topological order~\cite{Zeng2019Quantum} in 2D. Since nowadays such systems have become experimentally  accessible~\cite{Bloch2008Many-body,Cazalilla2011One}---turning the concept of quantum simulator~\cite{Buluta2009Science,Cirac2012Goals,Bernien2017Probing} into a reality---the interest in them has sparked far beyond the academic level. One dimensional systems, in particular, have been intensively investigated in nonequilibrium settings~\cite{Polkovnikov2011Colloquium,Gogolin2016Equilibration,Gring2012Relaxation, Schemmer2019Generalized}. On the one hand, they can  exhibit exotic behaviours such as relaxation to non-thermal states~\cite{Rigol2007Relaxation,Essler2016Quench,Ilievski2016Quasilocal}; on the other hand, there are interacting 1D models that allow for non-perturbative solutions, either 
exact~\cite{Ilievski2015Complete} or asymptotic~\cite{Castro-Alvaredo2016Emergent,Bertini2016Transport}, and, especially in quantum spin chains, numerical approaches based on tensor 
networks~\cite{Ran2020Tensor} are particularly efficient. 

Among the most remarkable nonequilibrium phenomena observed in 1D systems, we point up prethermalization~\cite{Moeckel2008Interaction,Rigol2009Breakdown,Bertini2015Prethermalization,Babadi2015Far,Langen2016Prethermalization,Abanin2017A,Alba2017Prethermalization,Reimann2019Typicality,Durnin2021Nonequilibrium} and prerelaxation~\cite{Fagotti2014On,Bertini2015pre-relaxation}. Such slow dynamics appear, for example, in strong coupling limits as a result of emergent symmetries~\cite{Fagotti2014On, zadnik2023slow}. 
One can then develop effective descriptions that remain valid for possibly large but still intermediate times, we mention for example the strong-weak duality that allows one to replace  strong-coupling Hamiltonians with weak-coupling ones~\cite{MacDonald1988tU}. In a quantum spin chain, if the strong-coupling model has an axial symmetry, the interplay between the conservation of a component of the total spin and the emergent conservation of the operator that is strongly coupled in the Hamiltonian could result in  constrained dynamics---see, e.g., Ref.~\cite{Zadnik2021The}---which have been recently studied especially in the context of Hilbert space fragmentation~\cite{Moudgalya2022Hilbert}.  This is an example of a physical situation in which (quasi)particles, at high enough density, can get stuck. When that happens, in analogy with classical physics, we shall say that the system exhibits ``quantum jamming''. To the best of our knowledge, the quantum version of jamming was  originally identified by mapping classical dissipative systems at finite temperature into  quantum systems at zero temperature~\cite{Biroli2008Theory}---see also  Refs~\cite{Nussinov2013Mapping,Artiaco2021Quantum}.
The interest in it had come from the idea that jamming 
might be linked to a new kind of phase transition~\cite{Biroli2007A}, akin to the more famous glass transition, which could be approached by varying a thermodynamic quantity such as the density. 

The key question we address here is which phenomenology should be expected close to quantum jamming when the system is  perturbed out of unstable equilibrium 
rather than driven from or to thermodynamic equilibrium. 
In the systems we consider, indeed, the space of jammed states is spanned by product states, which could be reproduced, for example, in optical lattices with ultracold atoms~\cite{Weitenberg2011Single}. 
This opens the door to investigations into the dynamical effects of local perturbations after having prepared  the system in a jammed state. 

Aiming at both generality and simplicity, 
we focus on a class of  quantum spin-$\frac{1}{2}$ chains with local Hamiltonians that exhibit quantum jamming due to the fact that spins up are not allowed to jump over other spins up and, if there are none of them along the way, only jumps by multiples of a given number $y$ are allowed. Specifically, the Hamiltonians are constructed (through sums and products) with the building blocks
\begin{equation}\label{eq:build_blocks}
T^{s,-s}_{\ell,\ell+y}=\sigma_\ell^{s}\prod_{j=1}^{y-1}\frac{1-\sigma_{\ell+j}^z}{2}\sigma_{\ell+y}^{-s}\, ,\qquad \sigma^z_\ell\, ,
\end{equation}
where $s=\pm 1$, $\sigma_\ell^{\pm}=(\sigma_\ell^x\pm i\sigma_\ell^y)/2$, and $\sigma_\ell^\alpha$ act like Pauli matrices on site $\ell$ and like the identity elsewhere.   
Although the building blocks 
could seem unnatural, the underlying constraints echo the blockade effect in chains of Rydberg atoms~\cite{Bernien2017Probing,doslic2023complexity}, and indeed similar terms can arise in effective descriptions of conventional systems, such as 
in the  strong anisotropy limit 
of the Heisenberg XXZ chain~\cite{Zadnik2021The}; interactions of that form have also been recently shown to trigger coexistence of phases in fermionic systems~\cite{Gotta2021Two}.
Beyond question is that constrained systems of that kind are attracting more and more attention~\cite{krajnik2023universal,Borsi2023Matrix}. 

The model with Hamiltonian $\sum_\ell T^{+-}_{\ell-1,\ell+1}+T^{-+}_{\ell-1,\ell+1}$, known as ``dual folded XXZ''~\cite{Zadnik2021The} or, following the terminology of Refs~\cite{Pozsgay2021Integrable,Borsi2023Matrix}, ``hard-rod deformation of XX with length $2$'', is a special case of an integrable system introduced by Bariev in 1991~\cite{Bariev1991Integrable}. Recently Ref.~\cite{Bidzhiev2022Macroscopic} pointed out that, in that model, a local unjamming perturbation in a jammed product state has macroscopic effects. Ref.~\cite{Zadnik2022Measurement} exhibited a similar result in a simplified setting where it was possible to map the dynamics of a single impurity over a jammed state into the dynamics of a single spin up in the XX model ($\sum_{\ell}\sigma_\ell^x\sigma_{\ell+1}^x+\sigma_\ell^y\sigma_{\ell+1}^y$). 
Such a transformation is an instance of a duality mapping, described in the following, that has the practical effect of removing the kinetic constraints.  
In the specific case of the dual folded XXZ, it translates trivial noninteracting dynamics in XX 
into unusual dynamics that incorporate the inhomogeneity of the underlying jammed state into effective  deformations of the space; but not only that. The effect of the perturbation reaches macroscopic scales, i.e., the perturbation indelibly affects local observables within a region with extent proportional to the time. 

We reconsider this nonequilibrium problem from a more general perspective and show that the phenomenology characterised by macroscopic effects from local perturbations, which we will compactly call MELP,  survives both integrable and non-integrable interactions, at least as long as the Hamiltonian has a symmetry under ``transmutation of particle species'' (to be compared with the ``inertness'' property of Ref.~\cite{krajnik2023universal}). 
We show that the state that builds up from a local unjamming perturbation is quantum at a macroscopic scale, setting quantum jamming apart from its classical counterpart. 
The non-symmetric case requires a separate treatment and the phenomenology turns out to strongly depend on  the inhomogeneity of the jammed state. 
Specifically, MELP can be observed even in the absence of that symmetry whenever the jammed state is (almost) homogeneous; the more disordered the sequence of species, however, the more hindered the spreading of the perturbation.

\section{A survey of the model}
The beauty of the constrained model that we propose is that a significant part of its phenomenology can be understood without specifying the details of the interactions. This will put us in the privileged position to discern the effects of quantum jamming from the  complex dynamical properties of the underlying systems, which can be described by essentially any spin-$\frac{1}{2}$ Hamiltonian with an axial symmetry. 

\subsection{From spins to particles, from positions to macropositions}
The building blocks in \eqref{eq:build_blocks} commute with the total spin in the $z$ direction $S^z=\frac{1}{2}\sum_\ell\sigma_\ell^z$, hence spins up can be reinterpreted as particles that are conserved in number. In addition, the building blocks preserve the sequence  of positions of spins up modulo $y$, therefore it is reasonable to distinguishing $y$ species of particles, characterised by their position modulo $y$. 
We can then separate the information about the species from the position $\ell$ by defining the \emph{macroposition} $\ell'$ as
\begin{equation}
\ell=y\ell'-b\, ,
\end{equation}
where $b\in \{0,1,\dots, y-1\}$ labels the particle species. The macroposition plays the role of the position for the particles, hence there can be up to $y$ particles in the same macrosite.

The symmetry under transmutation of species that we mentioned before is present when the interactions 
are blind to the species. This happens in particular when the Hamiltonian can be written in terms of the local averages over $y$ sites of the building blocks, $\frac{1}{y}\sum_{n=0}^{y-1} T_{\ell-n,\ell-n+y}$ and $\frac{1}{y}\sum_{n=0}^{y-1}\sigma_{\ell+n}^z$.

We note that a family of jammed excited states can be readily identified independently of the Hamiltonian: 
any state in which all consecutive particles have macrodistance not larger than $1$ (i.e., the spins up are at a distance not larger than $y$) is jammed.

\subsection{Multispecies particle-hole duality transformation}
Since the sequence of species $\underline b$ is conserved, it is convenient to split the representation of the state  in two parts, one of which stores just the information about the sequence $\underline b$ and hence is stationary. By removing the information about the species, particles become indistinguishable; let us associate each of them with a pseudospin down.  A pseudospin up represents instead the absence of a particle, so the number of pseudospins up between two pseudospins down should match the macrodistance (i.e., the difference between ordered macropositions) between the corresponding particles minus one. 
\begin{figure*}
\begin{minipage}{13.4cm}
\includegraphics[width=0.45\textwidth]{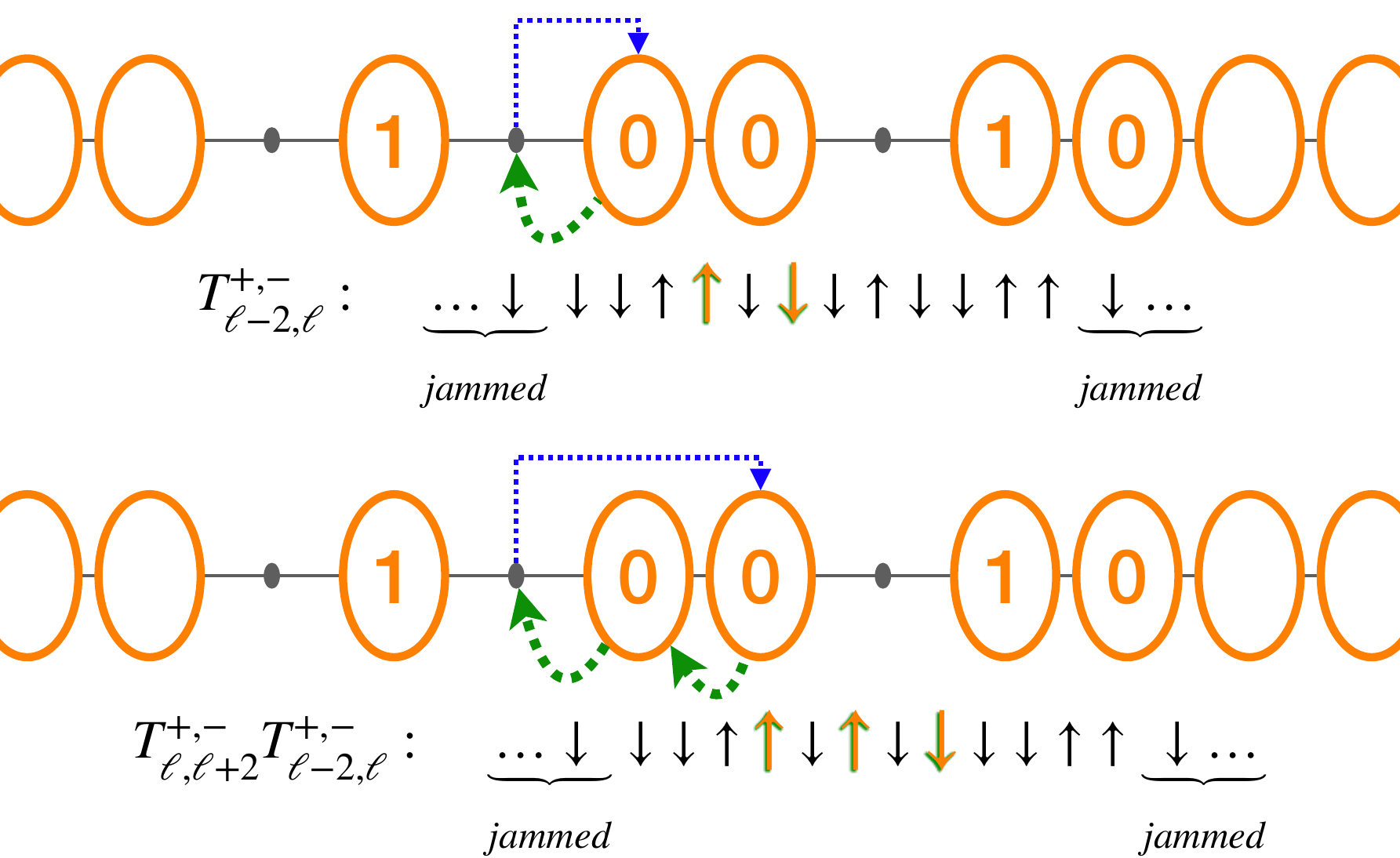}\hspace{0.3cm}
\includegraphics[width=0.45\textwidth]{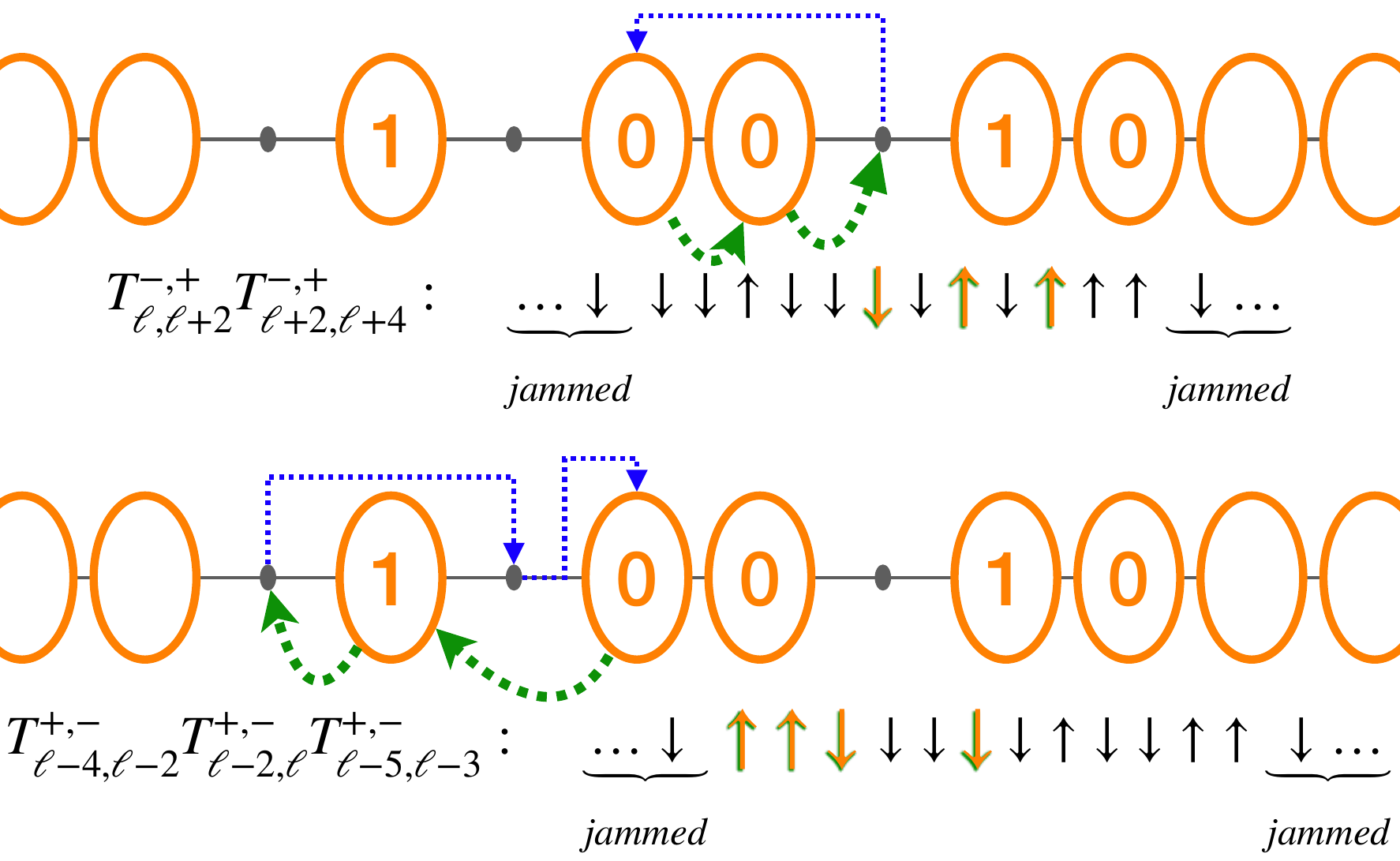}
\end{minipage}
\caption{
Cartoon of some actions allowed by the model with $y=2$, where  
orange ovals represent the (impenetrable) particles. The green (blue) arrows on the bottom (top) show the movement of the particles (impurities) in the exemplar state~\eqref{eq:examplestate}. The spins that are flipped (w.r.t.~\eqref{eq:examplestate}) are highlighted in orange.  
}\label{f:mapping}
\end{figure*}
We can interpret the mapping of spins into pseudospins for given sequence of species as a multispecies particle-hole duality transformation.

The prescription that we described could look plain at first sight but belies an important complication: macropositions, pseudopositions (i.e. the position of the pseudospins) and the relative position of a particle in the sequence of species $\underline b$ are not trivially related to one other. 
That is to say, there is arbitrariness in assigning the positions to a given spin, a given pseudospin, and a given particle, but, once the choice is made in a state, there is no freedom anymore and a link between the different positions is induced in any state.

For the purposes of this work, we can restrict ourselves to a sector with a fixed number $\nu$ of impurities (pseudospins up) over a jammed state. Let us then agree to assign macroposition $0$ to the particle with relative position $0$ in the state in which the impurities are at pseudopositions $0,1,\dots,\nu-1$; we also agree that, in that state, the particle with relative position $0$ corresponds to the pseudospin down at pseudoposition $-1$. The macroposition of the particle with relative position $\tilde j$ is then  given by
\begin{equation}\label{eq:ellp}
\ell'(\tilde j)=
\tilde j+\delta \tilde j+\sum_{k=1}^{\nu} \theta(n_k+1-k<\tilde j)
\end{equation}
where $n_1,\dots,n_\nu$ are the pseudopositions of the impurities  and $\delta \tilde j$ parametrises the deformation of the space due to the different distance that particles of different species have, 
\begin{equation}
\delta \tilde j=\sum_{\tilde n=0}^{|\tilde j|-1}\begin{cases}
-\theta_H(b_{\tilde n+1}-b_{\tilde n}< 0)&\tilde j\geq 0\\
\theta_H(b_{-\tilde n}-b_{-\tilde n-1}< 0)&\text{otherwise.}
\end{cases}
\end{equation}
An instance of such a deformation  has been already pointed out and discussed in Ref.~\cite{Zadnik2022Measurement} for  a single-impurity problem  with $y=2$. Since, to our present purposes, the deformation plays only a marginal role, we refer the reader to Ref.~\cite{Zadnik2022Measurement} for additional comments.

In the following we will denote the state with the sequence of species $\underline b$ and impurities at $n_1,\dots,n_\nu$ by $\ket{n_1,\dots,n_\nu;\underline b}$ (the absence of impurities will be denoted by  $\ket{\emptyset;\underline b}$). An example with $y=2$ follows
\begin{multline}\label{eq:examplestate}
\ket{\underbrace{\cdots \downarrow}_{jammed}\underbrace{\downarrow\downarrow}_{\ell'=-2}\underbrace{\uparrow\downarrow}_{-1}\underbrace{\downarrow\uparrow_{\ell=0}}_0\underbrace{\downarrow\uparrow}_1\underbrace{\downarrow\downarrow}_2\underbrace{\uparrow\uparrow}_3\underbrace{\downarrow\cdots}_{jammed}}\equiv \\
\ket{\Downarrow\uparrow\downarrow^{(1)}\uparrow\downarrow^{(0)}\downarrow_{j=0}^{(0)}\uparrow\downarrow^{(1)}\downarrow^{(0)}\Downarrow}\equiv \ket{-4,-2,1;\underline b}\, ,
\end{multline}
where $\underline b=\dots,1,0,0,1_{\tilde j=0},0,\dots$, $\Downarrow$ stands for a semi-infinite string of pseudospins down, and positions, macropositions, pseudopositions and relative positions are denoted by $\ell$, $\ell'$, $j$, and $\tilde j$, respectively;
the first representation in the second line shows the pseudospins  
with the species as superscripts; 
the last representation displays the state by indicating the pseudopositions of the pseudospins up followed by the sequence of species.  Fig.~\ref{f:mapping} shows how some allowed interactions act on \eqref{eq:examplestate}. 

For given $\underline b$, the dynamics can be projected onto the space of pseudospins by applying the duality transformation to the Hamiltonian. A procedure is described in Section~\ref{ss:mappingH}.
We find that symmetric translationally invariant  operators with local densities are represented in the pseudospace by  operators with densities that are still local and, in addition, are independent of the sequence of species. Non-symmetric operators, instead, have a sequence-dependent semilocal representation. Here we are using the term ``semilocal'' to indicate representations of local observables that do not commute with all local operators acting on the pseudospins at arbitrarily large pseudodistance; an example follows with \eqref{eq:HI}. 
\section{Results}
A qualitative understanding of MELP in the class of models we consider can be gained without specifying the details of the model. We do it focusing on the local magnetisation $\sigma_\ell^z$. This is an observable with diagonal form factors in the jammed sector, hence it is arguably the simplest operator we could study that witnesses MELP in a transparent way.   
We will then corroborate the qualitative analysis considering specific examples. 
\subsection{Local magnetisation}
Since the total spin in the $z$ direction is conserved, the local magnetisation is a charge density. As such, it satisfies a continuity equation, which typically admits of hydrodynamic descriptions. As pointed out in Ref.~\cite{Zadnik2022Measurement}, this kind of large-scale approaches are rather ineffective in the present case. Thus, we work out time evolution employing standard Lehmann representations of the correlations functions. We start here with the one-point function---the local magnetisation---whereas a two-point function will be discussed in Section~\ref{ss:macro}.
We introduce the notation 
\begin{equation}\label{eq:notation}
:O:=O-\braket{\emptyset;\underline b|O|\emptyset;\underline b}\, .
\end{equation}
In the basis $\ket{n_1,\dots,n_\nu;\underline b}$, the local magnetisation is a diagonal operator. Using~\eqref{eq:ellp}, we derive in Appendix~\ref{a:mag} the following matrix elements
\begin{multline}\label{eq:szn}
\braket{n;\underline b|:\sigma_{y\ell'-j}^z:|n; \underline b}=\\
\braket{\emptyset;\underline b|\sigma_{y(\ell'-1)-j}^z|\emptyset;\underline b}\theta_H(n+\delta n+1<\ell')\\
-\braket{\emptyset;\underline b|\sigma_{y\ell'-j}^z|\emptyset;\underline b}\theta_H(n+\delta n-\theta_H(b_{n+1}<b_n)< \ell')\\
+\delta_{\ell',n+\delta n+1}[\theta_H(b_{n+1}<b_n)\delta_{b_{n+1},j}-1]\\
+\delta_{\ell',n+\delta n}\bigl[\theta_H(b_{n+1}<b_{n}<b_{n-1})\delta_{b_{n-1},j}\\
-\theta_H(b_{n+1}<b_n)(1-\delta_{b_{n},j})\bigr]
\end{multline}
and
\begin{multline}\label{eq:sznnu}
\braket{n_1,n_2,\dots,n_{\nu};\underline b|:\sigma^z_{\ell}:|n_1,n_2,\dots,n_{\nu};\underline b}=\\
\sum_{k=1}^{\nu}\braket{n_k+1-k;\underline b|:\sigma_{\ell-y(k-1)}^z:|n_k+1-k;\underline b}
\end{multline}
Roughly speaking, the local magnetisation in a product state with a single impurity---\eqref{eq:szn}---is identified with either one local magnetisation or another ($y$ spins apart) in the underlying jammed state, depending on whether the impurity is on the left or on the right. The matrix elements with more impurities---\eqref{eq:sznnu}---have a similar interpretation and can be reduced to those with a single impurity. 

Time evolution is readily obtained 
\begin{equation}\label{eq:szt}
\braket{: \sigma^z_{\ell}:}=\sum_{k=1}^\nu\sum_n p_{n+k-1,t}^{(k)}\braket{n;\underline b|:\sigma_{\ell-y(k-1)}^z:|n;\underline b}
\end{equation}
where $p^{(j)}_{n,t}$ is the probability that the $j$-th pseudospin up is at pseudoposition $n$ at time $t$, and it is given by
\begin{equation}
 p^{(j)}_{n,t}=\sum_{n'}|\braket{
 \dots,n'_{j-1},n,n'_{j+1},\dots
 |e^{-i\tilde{ H} t}|\tilde \Psi_0}|^2.
\end{equation}
Here $\ket{\tilde \Psi_0}$ is the projection of the initial state onto the pseudospin space, $\tilde{H}$ is the projected Hamiltonian,
and the sum is over all   integers $n'_1,\dots,  n'_{j-1},n'_{j+1},\dots,n'_{\nu} $ satisfying $n_1'<\dots<n'_{j-1}<n$ and $n<n'_{j+1}<\dots<n'_\upsilon$. 

For finite $\nu$, the qualitative behaviour of the magnetisation can be readily understood if $\tilde H$ is local (i.e., the original Hamiltonian is symmetric) and doesn't exhibit exceptional  localisation properties in the eigenstates. It is then reasonable to expect $p^{(j)}_{n,t}$ to approach zero in time. Consequently, the terms in the last three lines of \eqref{eq:szn} give a subleading contribution to \eqref{eq:szt}, and the argument of the step functions on the second and third lines of  \eqref{eq:szn} can be shifted without affecting the asymptotic behaviour of the local magnetisation. We then end up with the following simplified expression
\begin{equation}\label{eq:sigmaasymp}
\braket{:\sigma^z_{\ell}:}\sim 
\sum_{k=1}^{\nu} 
\braket{\emptyset;\underline b|\sigma_{\ell-yk}^z-\sigma_{\ell-y (k-1)}^z|\emptyset;\underline b}F_{t}^{(k)}(x_\ell)
\end{equation}
where  
$F_{t}^{(k)}(x)=\sum_{n\lesssim x}p_{n,t}^{(k)}$ is the cumulative distribution and  $x_\ell$ is the effective pseudoposition associated with site $\ell$, i.e.,  
\begin{equation}\label{eq:x}
\frac{y}{\ell}(x_{\ell}+\delta x_{\ell})\sim 1+O(\ell^{-1})\, .
\end{equation}

For its generality and transparency, \eqref{eq:sigmaasymp} is the most important formula of this paper. 
\begin{itemize}
\item[-] Firstly, by modifying the underlying jammed state, the coefficients of $F_{t}^{(k)}(x_\ell)$ in \eqref{eq:sigmaasymp} (which belong to $\{-2,0,2\}$) can be changed in an arbitrary way without even affecting the asymptotics of $x_\ell$. 
This allows us to ignore the coefficients altogether and \emph{focus on the cumulative distributions.}
\item[-] Secondly, 
as clarified in the following examples, $p_{n,t}^{(k)}$ typically  decays as $1/t$ (at least for small enough $\nu$), and hence the typical variation of the cumulative distribution is over ballistic scales $x\sim t$. Thus, \emph{we predict a general nontrivial profile for the local magnetisation as a function of $x/t$}, which is the unusual lightcone behaviour pointed out in Refs~\cite{Bidzhiev2022Macroscopic,Zadnik2022Measurement} in a specific model with a single impurity.  
\item[-] Corrections should be expected if there is a finite number of \emph{localised eigenstates}, whose more explicit effect would be to \emph{create effective discontinuities in the relevant scaling limit of $F_t^{(k)}(x)$}.
\item[-]On the contrary, in order for the impurities \emph{not to change the local magnetisation in a region that grows in time} for a generic jammed product state, the probabilities  $p_{n,t}^{(k)}$ should remain localised around the original positions of the impurities, which is possible only if \emph{the sector with $\nu$ impurities is spanned by almost only localised excited states}. This is practically impossible when the Hamiltonian is symmetric and the interactions are local, but we will show in the examples below that it becomes an option in the presence of non-symmetric interactions. 
\end{itemize}
We would like to emphasise that these qualitative rules apply to the entire class of constrained models constructed with the building blocks in Eq.~\eqref{eq:build_blocks}, which include noninteracting, interacting integrable, and nonintegrable systems. 
\subsection{Example: symmetric Hamiltonian}
\begin{figure}[t]
\includegraphics[width=.4\textwidth]{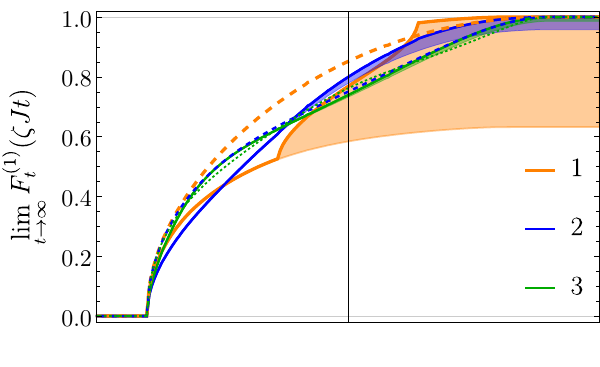}\\
\vspace{-0.5cm}
\includegraphics[width=.4\textwidth]{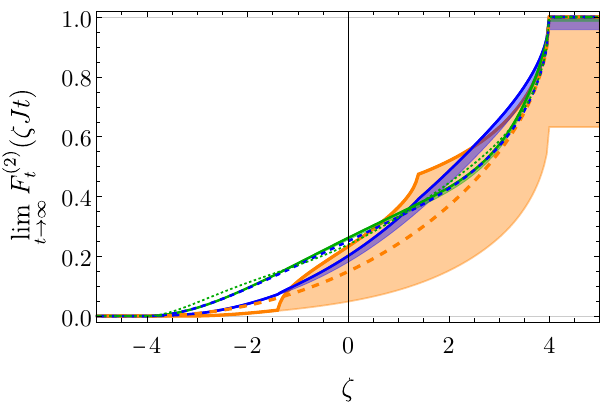}
\caption{The cumulative distributions for the (symmetric) example in \eqref{eq:H} with $\Delta=0.5$ and $g=-1$. Each solid curve corresponds to a different value of the pseudodistance $n_2^{(0)}-n_1^{(0)}$ between the impurities at the initial time. The associated dashed lines are obtained by trivialising the scattering phase. The region delimited by each curve represents the contribution from the bound states. 
}
\label{f:1}
\end{figure}
Ref.~\cite{Zadnik2022Measurement} has proven MELP in the special case in which the Hamiltonian is  local and noninteracting in the pseudospace.
In order to prove the robustness of that phenomenology under interacting and integrability breaking  terms, we consider the following Hamiltonian with $y=2$
\begin{multline}\label{eq:H}
H=J\sum\nolimits_\ell (\sigma_{\ell-1}^x \sigma_{\ell+1}^x+ \sigma_{\ell-1}^y \sigma_{\ell+1}^y+\Delta  \sigma_{\ell-1}^z \sigma_{\ell+1}^z)\tfrac{ 1- \sigma_\ell^z}{2}\\+g(\sigma_{\ell-1}^x\sigma_{\ell+3}^x+\sigma_{\ell-1}^y\sigma_{\ell+3}^y)\tfrac{1-\sigma_\ell^z}{2}\tfrac{1-\sigma_{\ell+1}^z}{2}\tfrac{1-\sigma_{\ell+2}^z}{2}
\end{multline}
The first line is a hard-rod deformation of the XXZ model and is  interacting integrable. 
The second line is a symmetric interaction that breaks integrability\footnote{As it is written, this Hamiltonian is symmetric only in the weak sense that the dependency on the species is a trivial constant fixed by the sequence $\underline b$, the exact symmetry requiring also to add an irrelevant constant and a coupling to $S^z$ (omitted here for the sake of compactness).}. The Hamiltonian of Ref.~\cite{Zadnik2022Measurement} is recovered by setting $\Delta=g=0$.
The projection of $H$ onto the pseudospace reads
\begin{multline}
\tilde H\sim J\sum\nolimits_{j} \tau_j^x \tau_{j+1}^x+ \tau_j^y \tau_{j+1}^y+\Delta  \tau_j^z \tau_{j+1}^z\\
+g (\tau^x_{j-1}\tau^x_{j+1}+\tau^y_{j-1}\tau^y_{j+1})
\tfrac{ 1+ \tau^z_j}{2}\, ,
\end{multline}
where we indicated the pseudospins operators by $\frac{1}{2}\vec \tau_\ell$ 
to avoid confusion with the spin operators. 
\subsubsection{Single impurity: noninteracting representation}
Quite generally, the physics of a single impurity can be described by an effective noninteracting Hamiltonian. Our example is no exception and, in fact, we find that the coupling constants $\Delta$ and $g$ that set this example apart from the problem studied in Ref.~\cite{Zadnik2022Measurement} are irrelevant. Specifically, 
the additional interactions results in just a shift of the excitation energy, given by $E(p)=4 J(\cos p-\Delta)$, that doesn't have any impact on the local magnetisation. 
\subsubsection{Two impurities: integrable representation}
Much more interesting is the case in which there are two impurities in the initial state. In this sector interactions start playing a role, however integrability is effectively preserved, in the sense that the Bethe Ansatz~\cite{Bethe1931Zur,Essler2005The} provides sufficient degrees of freedom to solve the problem. 
We obtain the scattering phase
\begin{equation}\label{eq:S}
S(p_1,p_2)\!=\!-\tfrac{1-2[\Delta +g\cos(p_1+p_2)]e^{i p_2}+e^{i (p_1+p_2)}}{1-2[\Delta+g\cos(p_1+p_2)]e^{i p_1}+e^{i (p_1+p_2)}}.
\end{equation} 
Importantly, the corresponding integrable model 
exhibits also a bound state in the 2-particle sector, provided that the momentum $p$ satisfies $1+\cos p<2(\Delta + g \cos p)^2$. The bound state consists of two impurities with total momentum $p$ travelling at a typical pseudodistance $1/\log|\frac{ \Delta +g  \cos p}{\cos(p/2)}|$ with energy $E_2(p)=4J(\cos^2(p/2)+g^2\cos^2p-\Delta^2)/(\Delta +g\cos  p)$.
The large time limit of $F_t^{(k)}(x)$ takes contributions from both magnons (m) and bound states (b), $F_t^{(k)}(x)=F_{t,\mathrm{m}}^{(k)}(x)+F_{t,\mathrm{b}}^{(k)}(x)$. They are worked out in Appendix~\ref{a:Bethe} and are given by
\begin{multline}
F_{t,\mathrm{m}}^{(j)}(x)\sim \int_{-\pi}^\pi\frac{\mathrm d^2 p}{8\pi^2}\theta_H(\tfrac{x}{t}-v^{(j)}(p_1,p_2))\\
|e^{i(n_1^{(0)} -n_2^{(0)})(p_1-p_2)}+S(p_1,p_2)|^2\, ,\label{eq:Fmagnon}
\end{multline}
\begin{multline}
F_{t,\mathrm{b}}^{(1)}(x)\sim F_{t,\mathrm{b}}^{(2)}(x)\sim \int_{-\pi}^\pi\frac{\mathrm d p}{2\pi}\theta_H(\tfrac{x}{t}-E_2'(p))\\
\max\bigl[\bigl(\tfrac{\Delta +g \cos  p}{\cos \frac{p}{2}}\big)^2-1,0\bigr] \bigl(\tfrac{\cos^2\frac{ p}{2}}{(\Delta +g  \cos p)^2}\bigr)^{n_2^{(0)}-n_1^{(0)}}\, ,\label{eq:Fbound}
\end{multline}
where 
$$
v^{(j)}(p_1,p_2)=\begin{cases}
\min(E'(p_1),E'(p_2))&j=1\\
\max(E'(p_1),E'(p_2))&j=2\, .
\end{cases}
$$
In the equations $n_1^{(0)}<n_2^{(0)}$ denote the pseudopositions of the impurities at the initial time
(note that we took the limit $x,t\rightarrow \infty$ for given $n_1^{(0)}, n_2^{(0)}$, so the impurities in the initial state correspond to $x\sim 0$). 

Figure~\ref{f:1}  highlights the effect of the interactions by comparing the cumulative distributions with those obtained by trivialising the scattering phase, i.e., \mbox{$S(p_1,p_2)\rightarrow -1$}. The bound state contributes significantly only when the impurities at the initial times are close enough. The dissimilarity between the maximal velocities of magnons and bound states  is manifested in the appearance of cusps within the lightcone. For the readers familiar with Bethe Ansatz, we specify that those can be interpreted as bare velocities, indeed the state differs from the reference state by a finite number of excitations. Note however that there is an implicit dressing coming from the deformation of the space, i.e., from the relation,~\eqref{eq:x}, between $x$ and $\ell$, which can be read as a simplified instance of the effects pointed out in Ref.~\cite{Bonnes2014Light}.
\subsection{Example: non-symmetric Hamiltonian}
The symmetry of the Hamiltonian shown in \eqref{eq:H} is broken by the interaction
\begin{equation}\label{eq:HI0}
H_I=J V\sum\nolimits_\ell \sigma_\ell^z\sigma_{\ell+1}^z\, ,
\end{equation}
which has the following semilocal representation in the pseudospace
\begin{multline}\label{eq:HI}
\tilde H_I=4JV\sum\nolimits_j \sum\nolimits_{\tilde j}\tfrac{1-(-1)^{b_{\tilde j}-b_{\tilde j+1}}}{2}\int_{-\pi}^\pi\frac{\mathrm d \omega}{2\pi} \\
 e^{i\omega (\tilde j-j-1+\sum_{j'=-\infty}^j\frac{ 1+\tau^z_j}{2})}\tfrac{ 1-\tau_j^z}{2}\tfrac{ 1-\tau_{j+1}^z}{2}\, .
\end{multline}
Here the integral over $\omega$ represents a Kronecker delta.  In the following we will consider time evolution under $H+H_I$.  In the specific case of a single impurity, the total Hamiltonian acting on the pseudospace can be represented by the following noninteracting operator
\begin{equation}\label{eq:Hnonsym1}
\tilde{H}_1\sim J\sum\nolimits_{j} \!\!\tau_j^x \tau_{j+1}^x+ \tau_j^y \tau_{j+1}^y\!-V(1-(-1)^{b_{j}-b_{j+1}})  \tau_j^z\, ,
\end{equation}
from which we removed an irrelevant term proportional to the $z$ component of the total pseudospin.
\subsubsection{Quasi-homogeneous jammed state}
The semilocality of the interaction is masked if the sequence of particles consists of a repeating pattern. In order to emphasize the differences with respect to the symmetric case, we study the same example considered in Ref.~\cite{Zadnik2022Measurement}: the  sequence of species of the underlying jammed state consists of the repeating pattern $\{1,0,0,1\}$ and the perturbation creates an impurity by removing a particle. We can fix the notations in such a way that the initial state is given by
$
\ket{\cdots
\uparrow\uparrow\downarrow\uparrow\uparrow\downarrow_0\downarrow\uparrow\downarrow\uparrow\uparrow\downarrow
\cdots }
\sim \ket{0;\underline b}
$, 
where the sequence of species reads
\begin{equation}\label{eq:b}
\underline b= \dots,(1,0,0,1),(1,0,0,1_0),(0,0,1,1),(0,0,1,1),\dots
\end{equation}
Here we displayed relative position $0$ and grouped the species to highlight the repeating pattern.
The almost periodicity of the species translates into the nearly two-site shift invariance of $\tilde H_1$ in \eqref{eq:Hnonsym1}, a symmetry that is broken around $j=0$, where the effective staggered spin coupled by $V$ changes sign. Generally, step potentials can give rise to localised excited states. This is confirmed by our numerical investigation, which  shows that the initial state has a significant overlap with a localised excited state. Nevertheless, there is a delocalised contribution that triggers MELP. This is shown in the first row of Figure~\ref{f:disorder}, where we can also see that the localised excited state creates a discontinuity in the cumulative distribution function (the sharp change of colour around $x=0$) that is more pronounced as $V$ is increased.

The problem becomes considerably more complicated when the initial state has two impurities, but it is still reasonable to expect an  integrable description almost everywhere, and hence MELP. We surmise however that the effects could become less pronounced due to the localised inhomogeneities of the interactions. 

\subsubsection{Disordered jammed state: localisation}
\begin{figure*}
\centering
\includegraphics[width=11.4cm]{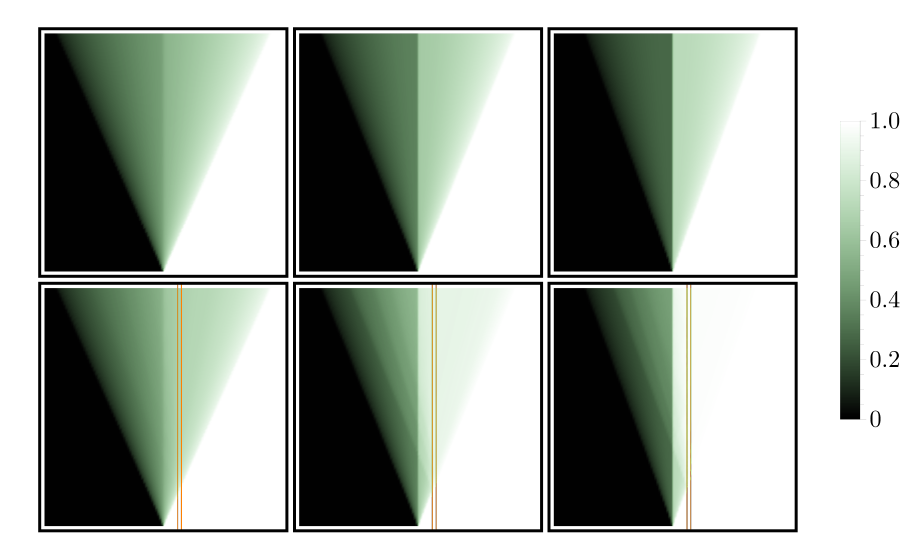}
\caption{The cumulative distribution $F_t^{(1)}(x)$ for a single impurity as a function of  pseudoposition ($x$ axis) and time $Jt$ ($y$ axis) in the (non-symmetric) example described by  \eqref{eq:Hnonsym1} starting from $\ket{0;\underline b}$ with the sequence in \eqref{eq:b}. Each column corresponds to a different value of $V=\{0.1,0.3,0.5\}$ (the values of $\Delta$ and $g$ are irrelevant). The first row is for the configuration without disorder. The second raw is for a configuration with a small disordered region delimited by the two orange vertical lines.
}
\label{f:disorder}
\end{figure*}

The picture presented in the previous example breaks down if we add a disordered region in the sequence of species.
This creates a region in the pseudospace in which the term coupled by $V$ in \eqref{eq:Hnonsym1} is disordered, which is known to trigger Anderson's localisation~\cite{Anderson1958Absence}. 
As shown in the second row of Figure~\ref{f:disorder}, 
for large enough $V$ the information about the impurity is not able to cross the barrier.  Thus, the jammed state on one side of the disorder barrier is protected against a localised perturbation on the opposite side, provided that the width of the barrier is large enough (with respect to $1/V^2$).
\subsection{Macroscopic quantum states}\label{ss:macro}
The problem under investigation shares some similarities with the MELP studied in Ref.~\cite{bocini2023growing}, which considered local perturbations in low-entangled excited states that cannot be interpreted as ground states of local conservation laws. 
Following Ref.~\cite{bocini2023growing} (see also Ref.~\cite{Bidzhiev2022Macroscopic}), let us then 
imagine to prepare the system in a jammed product state $\ket{\Psi(0^-)}$ (with spins along $\hat z$). After identifying a site $\ell_0$ with a spin up whose flip would unjam the state, we  perform a projective measurement of $\hat n\cdot \vec \sigma_{\ell_0}$, with $\hat n$ different from $\hat z$. 
The state after the measurement $\ket{\Psi(0^+)}$ will be in a superposition of the original jammed state and an almost identical quasi-jammed state with an impurity, $\ket{\Psi(0^+)}=\sqrt{p} \ket{\emptyset;\underline b}+e^{i\phi}\sqrt{1-p} \ket{n_0;\underline b^{(-)}}$ ($\phi$ and $p$ depend on $\hat n$), with probability $p$. 
By the Lieb-Robinson bound~\cite{Lieb1972The}, the speed at which the information about the impurity spreads is bounded from above, therefore the state remains separable outside a lightcone growing linearly in time. We restrict to the corresponding subsystem $\Omega_t$ (with extent proportional to the time, $|\Omega_t|\propto t$).
Let us then consider an extensive operator of the form $S_{\vec s}^z=\frac{1}{2}\sum_\ell s_\ell\sigma_\ell^z$, with $s_\ell=\pm 1$, which has the property not to fluctuate outside the lightcone. Because of that, its variance matches  the variance of its restriction to $\Omega_t$. 

The square root of the variance of such an operator provides a lower bound to the minimal system size for which, to use the words of Ref.~\cite{Frowis2012Measures}, ``one has to assume validity of quantum mechanics in order to explain a non-classical phenomenon''. Let us then compute the variance of $S_{\vec s}^z$. 
Denoting its expectation value and its variance at time $t$ by $\braket{S^z_{\vec s}}_t$ and $\braket{S^z_{\vec s},S^z_{\vec s}}_t$, respectively, we have 
\begin{multline}\label{eq:variance}
\braket{S^z_{\vec s},S^z_{\vec s}}_t\sim \frac{p}{1-p}(\braket{S^z_{\vec s}}_t-\braket{S^z_{\vec s}}_{0^-})^2\\
+(1-p)\braket{n_0;\underline b^{(-)}|e^{i H t}S^z_{\vec s},S^z_{\vec s}e^{-i H t}|n_0;\underline b^{(-)}}\, .
\end{multline}
Assuming \eqref{eq:sigmaasymp}, the right hand side of this equation grows as $t^2$ because the difference in the expectation values grows linearly in time. In fact, we are about to show that, at least in the presence of the symmetry under transmutation of species, 
the same conclusion applies also to the marginal case with $p=0$, which is the basic problem considered outside this section. 
The variance in the second line of \eqref{eq:variance} can indeed be expressed as follows
\begin{multline}
\braket{n_0|S_{\vec s}^z(t),S_{\vec s}^z (t)|n_0}=
\frac{1}{4}\sum_\ell (1-\braket{n_0|\sigma_\ell^z(t)|n_0}^2)+\\
\frac{1}{4}\sum_{\ell_1\neq \ell_2}s_{\ell_1} s_{\ell_2}\Bigl[\sum_{n}p_{n,t}(1-p_{n,t})\braket{n|:\sigma_{\ell_1}^z::\sigma_{\ell_2}^z:|n}-\\
\sum_{n\neq n'}p_{n,t}p_{n',t}\braket{n|:\sigma_{\ell_1}^z:|n}\braket{n'|:\sigma_{\ell_2}^z:|n'}\Bigr]\, ,
\end{multline}
where the dependency on $\underline b^{(-)}$ is understood and the last two lines depend on $n_0$ through $p_{n,t}$. The first term on the right hand side describes a standard extensive contribution to the variance; the interesting behaviour comes from the remaining terms. 
As discussed before, when the Hamiltonian is symmetric we can generally use the simplified expression
\begin{equation}
\braket{n;\underline b|:\sigma_{\ell}^z:|n; \underline b}\sim 
\braket{\emptyset;\underline b|\sigma_{\ell-y}^z-\sigma_{\ell}^z|\emptyset;\underline b}\theta_H(n<x_{\ell})
\end{equation}
which results in
\begin{multline}
\braket{S_{\vec s}^z,S_{\vec s}^z}_t\sim \frac{1}{4}\sum_{\ell_1\neq \ell_2}[F_{t}(\min(x_{\ell_1},x_{\ell_2}))-F_{t}(x_{\ell_1})F_{t}(x_{\ell_2})]\\
s_{\ell_1} s_{\ell_2}\braket{\emptyset;\underline b|\sigma_{\ell_1-y}^z-\sigma_{\ell_1}^z|\emptyset;\underline b}
\braket{\emptyset;\underline b|\sigma_{\ell_2-y}^z-\sigma_{\ell_2}^z|\emptyset;\underline b}\, .
\end{multline}
Let us then choose $s_\ell=\braket{\emptyset;\underline b|\sigma_\ell^z|\emptyset;\underline b}$; we find
\begin{multline}
\braket{S_{\vec s}^z,S_{\vec s}^z}_t\sim \sum_{\ell_1\neq \ell_2}[F_{t}(\min(x_{\ell_1},x_{\ell_2}))-F_{t}(x_{\ell_1})F_{t}(x_{\ell_2})]\\
\braket{\emptyset;\underline b|\tfrac{1-\sigma_{\ell_1-y}^z\sigma_{\ell_1}^z}{2}|\emptyset;\underline b}
\braket{\emptyset;\underline b|\tfrac{1-\sigma_{\ell_2-y}^z\sigma_{\ell_2}^z}{2}|\emptyset;\underline b}\, .
\end{multline}
Generally this grows as $t^2$, i.e., $|\Omega_t|^2$, indeed $F_t(x)$ approaches a function of $x/t$ and $\braket{\emptyset;\underline b|\frac{1-\sigma_{\ell-y}^z\sigma_{\ell}^z}{2}|\emptyset;\underline b}$ has a strictly positive local average. This implies that also the second line of \eqref{eq:variance}
grows quadratically in time.

We can therefore conclude that 
the local perturbation brings $\Omega_t$ to a macroscopic quantum state. That is to say, the MELP under investigation goes beyond classical physics and should be regarded as a signature of quantum jamming. 

\section{Discussion}
Macroscopic effects from local perturbations have been recently observed in several settings: ground states in ordered phases with spontaneous symmetry breaking~\cite{Zauner2015Time,Eisler2020Front}, symmetric initial states in the presence of semilocal conservation laws~\cite{Fagotti2022Global,fagotti2022nonequilibrium}, 
separable quantum scars~\cite{bocini2023growing}, 
jammed states in constrained systems~\cite{Bidzhiev2022Macroscopic,Zadnik2022Measurement}. 
This work falls within the last category and has lifted the observations of Refs~\cite{Bidzhiev2022Macroscopic,Zadnik2022Measurement} to a new level of generality: We have analytically studied the robustness of MELP under the inclusion of interactions, no matter if integrable or not; we have identified the key role played by a symmetry and the change in the phenomenology due to its explicit breaking; we have shown that the described phenomenon is genuinely quantum.
This was discussed in full generality within a class of spin-chain models that allow for a particle description and exhibit a sector of eigenstates spanned by jammed product states. To explicitly show the most important features identified, we have  worked out the dynamics in simple examples.

Quantum jamming appears as a phenomenon that mixes up microscopic with macroscopic,  classical with quantum. 
We discussed a duality transformation that, when looked over in the opposite direction, 
maps conventional local $U(1)$ symmetric models into constrained local models with jammed sectors, and it brings, in turn, aspects that are typically microscopic, such as scattering phases and formation of bound states, to macroscopic scales. We have seen in particular that, in a constrained system,  the expectation value of the local magnetisation---which is linearly related to its probability distribution---becomes, in the dual unconstrained system, the cumulative distribution. 

A question that remains open is about the emergence of a diffusive MELP, in which the effect spreads out as $\sqrt{t}$. This was reported in Ref.~\cite{Bidzhiev2022Macroscopic}  starting from a quasi-jammed product state with a single impurity ($\nu=1$) but a fluctuating sequence of species. 
A related question that we have not addressed is about the first signatures of integrability breaking, which we expect 
in the sector with three impurities. 

In the integrable case it is instead still unclear the relationship between the lightcone dynamics emerging because of quantum jamming and those emerging because of the existence of infinitely many conservation laws, which have become the signature of the theory of generalised hydrodynamics. 

\section{Methods}
\subsection{Explicit construction of the Hamiltonian in the example}
We clarify here why the Hamiltonians in the examples belong to the class of models characterised by the building blocks in \eqref{eq:build_blocks}. 
Specifically, it turns out that those Hamiltonians have both a linear and a quadratic dependency on the building blocks: up to a constant and a term proportional to $S^z$, which is conserved and irrelevant, the Hamiltonian in \eqref{eq:H} reads
\begin{multline*}
H\sim 2J\sum\nolimits_\ell  \bar T^{+-}_{\ell,\ell+2}+ \bar T^{-+}_{\ell,\ell+2}+2\Delta\{\bar T^{+-}_{\ell,\ell+3},\bar T^{-+}_{\ell,\ell+3}\}\\
+g (T_{\ell,\ell+2}^{\mp}  T_{\ell-2,\ell}^{\mp}+T_{\ell-2,\ell}^{\pm}  T_{\ell,\ell+2}^{\pm})\, ,
\end{multline*}
where we expressed in particular the interaction coupled by $\Delta$ in terms of the averaged building blocks, 
$
\bar T^{+-}_{\ell-1,\ell+2}=(T_{\ell-1,\ell+1}+T_{\ell,\ell+2})/2
$, 
so as to make it explicit the symmetry under transmutation of particles. The symmetry of the last term can be understood as it describes a double hopping in the same direction, completely independent of the parity of the spin up that hops.  

The additional interactions that we added to break the symmetry under transmutation of species do not need additional comments because they were 
already expressed in terms of the building blocks $\sigma_\ell^z$. 

\subsection{Duality mapping of shift invariant operators}\label{ss:mappingH}
A key ingredient to solve the dynamics in the class of constrained models that we considered is the Hamiltonian acting on the space of pseudospins. 
Using translational invariance, an effective procedure to determine the representation of one-site shift invariant operators in the pseudospace is to inspect their action when applied to fragments of a product state that are sufficient to reconstruct the state and that are mapped into elementary fragments in the pseudospace. For example, the action of the identity (which is trivial but not simply represented), of $\sum_\ell\sigma_\ell^z$ (which is as  trivial as the identity because it is conserved), and of $\sum_\ell\sigma_\ell^z\prod_{j=1}^{m-1}\frac{1-\sigma_{\ell+j}}{2}\sigma_{\ell+m}^z$ for $m\geq 1$ can by inferred by applying them to a sequence of spins starting with a spin up and ending before the next spin up, 
$\ket{\dots \uparrow_j\underbrace{\downarrow\dots\downarrow}_n(\uparrow_{j+n+1})\dots}
$. 
Such a fragment is indeed mapped into the elementary fragment
$
\ket{\dots\downarrow^b \underbrace{\uparrow\dots\uparrow}_{\lfloor n/y\rfloor}(\downarrow^{b-n-1\!\!\!\pmod y})\dots}
$,
where arrows represent here pseudospins. It is then easy to identify the pseudospin operator that acts on the dual fragment in the same way as the original spin operator does on the original fragment.  
This is explicitly done in Appendix~\ref{a:mapop}.
\subsection{Bethe Ansatz with two impurities}
In order to solve the two-impurity sector of the Hamiltonian in \eqref{eq:H}  we have used the  Bethe Ansatz
\begin{equation*}
\sum_{n_1<n_2} [e^{in_1 p_1+in_2 p_2}+S(p_1,p_2)e^{in_1 p_2+in_2 p_1}]\ket{n_1,n_2;\underline b}\, .
\end{equation*}
In the symmetric example we obtain the scattering phase in \eqref{eq:S} and energy $E(p)=4J(\cos p-\Delta)$.
By applying an artificial finite-volume regularisation with periodic boundary conditions, we find that the effective integrable model with scattering phase $S(p_1,p_2)$ exhibits also bound states in the 2-particle sector.  The corresponding excited states can be parametrised as follows
\begin{equation*}
\sum_{n_1<n_2}e^{i\frac{n_1+n_2}{2}p} \Bigl(\frac{\cos\frac{ p}{2}}{ \Delta +g  \cos p}\Bigr)^{n_2-n_1}\ket{n_1,n_2;\underline b}
\end{equation*}
provided that $1+\cos p<2(\Delta + g \cos p)^2$. Their energy reads $E_2(p)=4J(\cos^2\frac{ p}{2}+g^2\cos^2p-\Delta^2)/(\Delta +g\cos  p)$. The reader can find more details in Appendix~\ref{a:Bethe}, which also includes the derivation of \eqref{eq:Fmagnon} and \eqref{eq:Fbound}  
\begin{acknowledgments}
I am grateful to Saverio Bocini for sharing some data of tDMRG simulations (not included here); I also thank him for useful discussions.
This work was supported by the European Research Council under the Starting Grant No. 805252 LoCoMacro. 
\end{acknowledgments}

\newpage

\appendix
\onecolumngrid
\section{Expectation values of the local magnetisation}\label{a:mag}
Equation~\eqref{eq:ellp} links the macroposition of the spins up with their relative position in the sequence. 
Since, in a product state of spins aligned along $\hat z$, $\braket{\frac{1+\sigma_\ell^z}{2}}$ is nonzero and equal to $1$ only if there is $\tilde j$ such that $\ell=2\ell'(\tilde j)-b_{\tilde j}$,  we have
\begin{equation}\label{eq:szsemi}
\sigma_\ell^z= \sigma_{2\lceil\frac{\ell}{y}\rceil-[(-\ell)\!\!\!\pmod y]}^z\rightarrow -1+2\sum_{\tilde j}\delta_{b_{\tilde j},(-\ell)\!\!\!\pmod y}\int\frac{\mathrm d \tilde \omega}{2\pi} e^{i\tilde \omega(\ell'(\tilde j)-\lceil\frac{\ell}{y}\rceil)}
\end{equation}
This expression can be simplified at a fixed number of impurities, as will be done in the following. 
\subsubsection{Zero impurities: jammed states}
In a jammed state (which is the vacuum $\ket{\emptyset}$ in the pseudospace) we have
\begin{equation}
\ell'(\tilde j)=\tilde j+\delta j\, .
\end{equation}
Thus, the local magnetisation takes the following value
\begin{equation}\label{eq:sz_0}
\braket{\emptyset|\sigma_{y\ell'-j}^z|\emptyset}=-1+2\sum_{\tilde j}\delta_{\ell',\tilde j+\delta\tilde j}\delta_{b_{\tilde j},j}\, .
\end{equation}
\subsubsection{One impurity}
We denote by $\ket{n}$ the state with an impurity at position $n$ in the pseuodspace. We have 
\begin{equation}
\ell'(\tilde j)=\tilde j
+\delta \tilde j+\theta_H(\tilde j>  n)
\end{equation}
and hence
\begin{equation}\label{eq:sz_n}
\braket{n|\sigma_{y\ell'-j}^z|n}=-1+2\sum_{\tilde j\in B}\delta_{b_{\tilde j},j}\delta_{\tilde j+\delta\tilde j+\theta_H(\tilde j>n),\ell'}\, .
\end{equation}
We are going to express these matrix elements in terms of those in the underlying jammed state using \eqref{eq:sz_0}. 
We start from the identities
\begin{eqnarray}
\sum_{\tilde j>n}\delta_{b_{\tilde j},j}\delta_{\tilde j+\delta\tilde j+1,\ell'}=\frac{1+\braket{\emptyset|\sigma_{y(\ell'-1)-j}^z|\emptyset}}{2}-\sum_{\tilde j\leq n}\delta_{b_{\tilde j},j}\delta_{\tilde j+\delta\tilde j+1,\ell'}\\
\sum_{\tilde j\leq n}\delta_{b_{\tilde j},j}\delta_{\tilde j+\delta\tilde j,\ell'}=\frac{1+\braket{\emptyset|\sigma_{y\ell'-j}^z|\emptyset}}{2}-\sum_{\tilde j>n}\delta_{b_{\tilde j},j}\delta_{\tilde j+\delta\tilde j,\ell'}
\end{eqnarray}
Since $\tilde j+\delta \tilde j$ is a non-decreasing function of $\tilde j$, the sum on the right hand side of the first equation vanishes if $n+\delta n+1<\ell'$. On the other hand, the sum on the left hand side of the first equation vanishes when $n+1+\delta(n+1)+1>\ell'$. An analogous argument applies to the second equation. Using
$$
\delta(n+1)-\delta n=-\theta_H(b_{n+1}<b_n)
$$
we finally get
\begin{eqnarray}
\sum_{\tilde j>n}\delta_{b_{\tilde j},j}\delta_{\tilde j+\delta\tilde j+1,\ell'}=\begin{cases}\frac{1+\braket{\emptyset|\sigma_{y(\ell'-1)-j}^z|\emptyset}}{2}& n+\delta n+1<\ell'\\
0&n+\delta n+1-\theta_H(b_{n+1}<b_n)\geq \ell'\\
\delta_{b_{n+1},j}&b_{n+1}<b_{n}, \ell'=n+\delta n+1
\end{cases}\\
\sum_{\tilde j\leq n}\delta_{b_{\tilde j},j}\delta_{\tilde j+\delta\tilde j,\ell'}=\begin{cases}\frac{1+\braket{\emptyset|\sigma_{y\ell'-j}^z|\emptyset}}{2}& n+\delta n-\theta_H(b_{n+1}<b_n)\geq \ell'\\
0&n+\delta n<\ell'\\
\theta_H(b_{n}<b_{n-1})\delta_{b_{n-1},j}+\delta_{b_{n},j}&b_{n+1}<b_n, \ell'=n+\delta n\end{cases}
\end{eqnarray}
and hence
\begin{multline}\label{eq:matrix_n}
\braket{n|\sigma_{y\ell'-j}^z|n}-\braket{\emptyset|\sigma_{y\ell'-j}^z|\emptyset}=\\
\braket{\emptyset|\sigma_{y(\ell'-1)-j}^z|\emptyset}\theta_H(n+\delta n+1<\ell')-\braket{\emptyset|\sigma_{y\ell'-j}^z|\emptyset}\theta_H(n+\delta n-\theta_H(b_{n+1}<b_n)< \ell')\\
 +\delta_{\ell',n+\delta n+1}[\theta_H(b_{n+1}<b_n)\delta_{b_{n+1},j}-1]
+\delta_{\ell',n+\delta n}[\theta_H(b_{n+1}<b_{n}<b_{n-1})\delta_{b_{n-1},j}-\theta_H(b_{n+1}<b_n)(1-\delta_{b_{n},j})]\, .
\end{multline}
\subsubsection{More impurities:  reduction to the single impurity case} 
From \eqref{eq:ellp} and \eqref{eq:szsemi} it follows
\begin{multline}
\braket{n_1,n_2,\dots,n_{\nu}|\sigma^z_{y \ell'-j}|n_1,n_2,\dots,n_{\nu}}=
\braket{\emptyset|\sigma_{y\ell'-j}^z|\emptyset}-2\sum_{\tilde j\in B}\delta_{b_{\tilde j},j}\delta_{\tilde j+\delta j,\ell'}+2\sum_{\tilde j\in B}\delta_{b_{\tilde j},j}\delta_{\tilde j+\delta \tilde j+\sum_{k=1}^{\nu} \theta(n_k+1-k<\tilde j),\ell'}=\\
\braket{\emptyset|\sigma_{y\ell'-j}^z|\emptyset}+2\sum_{k=1}^{\nu}\sum_{\tilde j>n_k+1-k}\delta_{b_{\tilde j},j}(\delta_{\tilde j+\delta \tilde j,\ell'-k}-\delta_{\tilde j+\delta \tilde j,\ell'-k+1})
\end{multline}
We can then use
$$
\braket{n|\sigma_{y\ell'-j}^z|n}-
\braket{\emptyset|\sigma_{y\ell'-j}^z|\emptyset}=2\sum_{\tilde j>n}\delta_{b_{\tilde j},j}(\delta_{\tilde j+\delta\tilde j,\ell'-1}-\delta_{\tilde j+\delta\tilde j,\ell'})
$$
to recast the expectation value as follows
\begin{equation}\label{eq:reduction}
\braket{n_1,n_2,\dots,n_{\nu}|:\sigma^z_{\ell}:|n_1,n_2,\dots,n_{\nu}}=
\sum_{k=1}^{\nu}\braket{n_k+1-k|:\sigma_{\ell-y(k-1)}^z:|n_k+1-k}
\end{equation}
where we used the notation introduced  in Eq.~[5] of the  in text, i.e., 
\begin{equation}\label{eq:notation}
:O:=O-\braket{\emptyset;\underline b|O|\emptyset;\underline b}\, .
\end{equation}
Note that a similar decomposition holds in standard $U(1)$ symmetric models in which $n_j$ simply represents the position of the $j$-th spin up; the main difference is that here both the position of the operator and the pseudopositions of the impurities are shifted depending on the number of the impurities on their left.  

\section{Mapping of diagonal operators}\label{a:mapop}
In this section  we discuss the mapping of diagonal operators (in a basis diagonalising $\sigma^z$) onto the pseudospace. 
It is convenient to split them in classes depending on the minimal number of consecutive domains of spins down that are required to characterise their local action. 
\subsection{Class I}
The simplest operators are determined by their action on each fragment of a product state of the form 
\begin{equation}
\ket{\dots \uparrow_j\underbrace{\downarrow\dots\downarrow}_n(\uparrow_{j+n+1})\dots}\, .
\end{equation}
The corresponding contribution to their eigenvalue is summarised in Table~\ref{t:1}.
\begin{table*}[!htbp]
\centering
\begin{tabular}{c|c}
&$\ket{\dots \uparrow_j\underbrace{\downarrow\dots\downarrow}_n(\uparrow_{j+n+1})\dots}$\\
\hline
$1$&$n+1$\\
$\sigma^z$&$1-n$\\
$\sigma^z\otimes \sigma^z$&$n-3+4\delta_{n,0}$\\
$ \sigma^z\otimes\frac{ 1-\sigma^z}{2}\otimes \sigma^z$&$n-4+4(\delta_{n, 0}+\delta_{n,1})$\\
$ \sigma^z\otimes (\frac{ 1-\sigma^z}{2})^{\otimes 2}\otimes  \sigma^z$&$n-5+4(\delta_{n,1}+\delta_{n,2})+5\delta_{n,0}$\\
$ \sigma^z\otimes (\frac{ 1-\sigma^z}{2})^{\otimes 3}\otimes  \sigma^z$&$n-6+4(\delta_{n,3}+\delta_{n,2})+5\delta_{n,1}+6\delta_{n,0}$
\end{tabular}\caption{Local contribution to the eigenvalues of some class I operators ordered by their range.}\label{t:1}
\end{table*}
The duality mapping (from spins to pseudospins) acts as follows
\begin{equation}
\ket{\dots \uparrow_j\underbrace{\downarrow\dots\downarrow}_n(\uparrow_{j+n+1})\dots}\mapsto \ket{\dots\downarrow^b \underbrace{\uparrow\dots\uparrow}_{\lfloor \frac{n}{y}\rfloor}(\downarrow^{b-n-1\!\!\!\pmod y})\dots}\, ,
\end{equation}
where, on the right hand side, the superscripts of the down arrows label the species of particles, which we called $b$ for the first spin down. On the contrary, if the image of the state is $\ket{\dots\downarrow^{b_{\tilde j}} \underbrace{\uparrow\dots\uparrow}_{m}(\downarrow^{b_{\tilde j+1}})\dots}$, the number $n$ of spins down in the original fragment~is
\begin{equation}\label{eq:ny}
n=y m+(b_{\tilde j}-b_{\tilde j+1}-1)\!\!\! \pmod y=y m+\frac{y-1}{2}
-\frac{1}{2}\sum_{n=1}^{y-1}\frac{\sin(\frac{2\pi n (b_{\tilde j}-b_{\tilde j+1}-\frac{1}{2})}{y})}{\sin(\frac{n\pi}{y})}\, ,
\end{equation}
where we used
\begin{equation}
x\!\!\!\pmod y=\frac{y-1}{2}-\frac{1}{2}\sum_{n=1}^{y-1}\frac{\sin(\frac{2\pi n (x+\frac{1}{2})}{y})}{\sin(\frac{n\pi}{y})}\, .
\end{equation}
\begin{table*}[!ht]
\centering
\begin{tabular}{c|c}
&$\ket{\dots\downarrow^b \underbrace{\uparrow\dots\uparrow}_{\lfloor \frac{n}{y}\rfloor}(\downarrow^{b-n-1\!\!\!\pmod y})\dots}$\\
\hline
$\frac{ 1- \tau^z}{2}$&$1$\\
$\frac{ 1+ \tau^z}{2}$&$\lfloor\frac{n}{y}\rfloor$\\
$\frac{ 1- \tau^z}{2}\otimes\bigl(\frac{ 1+ \tau^z}{2}\bigr)^{\otimes k}\otimes \frac{ 1- \tau^z}{2}$&$\delta_{\lfloor\frac{n}{y}\rfloor,k}$
\end{tabular}\caption{Local contribution to the eigenvalues of some diagonal operators in the pseudospace.}\label{t:2}
\end{table*}

The mapping of the operators in Table~\ref{t:1} can be worked out using \eqref{eq:ny} and the action of the  diagonal operators in the pseudospace reported in Table~\ref{t:2}. We recognise three kinds of contributions to the corresponding eigenvalues:
\begin{description}
\item[Additive constant] The contribution of an additive constant $c$ in the expressions appearing in the second column of Table~\ref{t:1} is given by the number of spins up multiplied by $c$. This is represented in the pseudospace by 
\begin{equation}
c\rightarrow c\sum_{j}\frac{ 1- \tau_j^z}{2}\, .
\end{equation}
In fact, the total pseudospin in the $z$ direction is conserved and depends only on the sequence of species. In the following we label by $B(\sim \mathbb Z)$ the set of the relative positions and call ``configuration space'' the associated space. 
In order to reduce confusion, we instead call $I(\sim \mathbb Z)$ the set of the pseudopositions.  
Using these notations, we see that the additive constant 
can also be represented in the configuration space as follows
\begin{equation}
c\rightarrow  c\sum_{\tilde j\in B}1\, ,
\end{equation}
which we prefer as it shows more explicitly  that constants are conserved. In the following we will always replace $\sum_{j\in I} \tau_j^z$ by $\sum_{j\in I}1-2\sum_{\tilde j\in B}1$.
\item[Number ($n$) of spins down] This contribution can be worked out using \eqref{eq:ny}, which links the number $n$ of spins down in between spins up with the number $\lfloor\frac{n}{y}\rfloor$ of pseudospins up in the corresponding domain in between pseudospins down.  In particular, there is also a term written solely in terms of the species, which can be treated as a constant by virtue of the conservation of the configuration. Putting all together we get
\begin{equation}
n\rightarrow \sum_{j\in I}y  1-\frac{y+1}{2}\sum_{\tilde j\in B}1-\frac{1}{2}\sum_{\tilde j\in B}\sum_{n=1}^{y-1}\frac{\sin(\frac{2\pi n (b_{\tilde j}-b_{\tilde j+1}-\frac{1}{2})}{y})}{\sin(\frac{n\pi}{y})}
\end{equation}
\item[Kronecker deltas $\delta_{n,k}$] These terms characterise the eigenvalues of the operators with densities that consist of the tensor product of $k$ projectors $\frac{ 1- \sigma^z}{2}$ onto spin down, preceded and followed by the projector $\frac{ 1+ \sigma^z}{2}$ onto spin up (cf. Table~\ref{t:2} for the analogue in the pseudospin space). The representation in the pseudospin space can then be derived from the identity
\begin{equation}\label{eq:deltan}
\delta_{n,k}=\delta_{\lfloor\frac{n}{y}\rfloor, \lfloor\frac{k}{y}\rfloor}\delta_{(b_{\tilde j}-b_{\tilde j+1}-1-k)\!\!\! \pmod y,0}
=\delta_{\lfloor\frac{n}{y}\rfloor, \lfloor\frac{k}{y}\rfloor}\frac{1}{y}\sum_{m=0}^{y-1} e^{2\pi i m\frac{b_{\tilde j}-b_{\tilde j+1}-1-k}{y}}
\end{equation}
which results in the representation
\begin{equation}\label{eq:deltatoT}
\delta_{n,k}\rightarrow \sum_{j\in I}( Z^{(1+\lfloor\frac{k}{y}\rfloor)}_{1+k\!\!\!\!\!\pmod y})_j\, ,
\end{equation}
where $( Z^{(n)}_{s\!\!\pmod y})_j$ is the following semilocal operator
\begin{equation}\label{eq:T}
( Z^{(n)}_{s\!\!\!\!\!\pmod y})_j=\int_{-\pi}^\pi\frac{\mathrm d \omega}{2\pi}\sum_{k\in B} \sum_{m=0}^{y-1} \frac{e^{2\pi i m\frac{b_{k}-b_{k+1}-s}{y}}}{y}
e^{i\omega (k-1-j+\sum_{j'}^j\frac{ 1+\tau^z_j}{2})}\frac{ 1-\tau_j^z}{2}\Bigl(\prod_{m=1}^{n-1}\frac{ 1+\tau_{j+m}^z}{2}\Bigr)\frac{ 1-\tau_{j+n}^z}{2}\, .
\end{equation}
Here the integral over $\omega$ represents a Kronecker delta and the sum over $m$ a condition on the species of particles modulo $y$. 
\end{description}
Before considering other classes of diagonal operators, we point out the identity
\begin{equation}\label{eq:sumT}
\sum_{s=1}^y( Z_{s\!\!\!\!\!\pmod y}^{(n)})_j=\frac{ 1-\tau_j^z}{2}\Bigl(\prod_{m=1}^{n-1}\frac{ 1+\tau_{j+m}^z}{2}\Bigr)\frac{ 1-\tau_{j+n}^z}{2}
\end{equation}
which shows that the sum over $s$ is a (strictly) local operator.
This allows us to identify some diagonal local operators that remain local after the duality transformation. They are obtained by summing \eqref{eq:deltatoT} over $k$ for $m y\leq k<(m+1)y$ and given integer $m$; that is to say,  operators
\begin{equation}
 V_{m}=\sum_\ell\sum_{n=my+1}^{m(y+1)} P^{(+)}_\ell \Bigl(\prod_{j=1}^{n-1} P^{(-)}_{\ell+j}\Bigr)  P^{(+)}_{\ell+n}\, ,
\end{equation}
where 
\begin{equation}
 P^{(\pm)}=\frac{ 1\pm  \sigma^z}{2}\, ,
\end{equation}
are symmetric under transmutation of species. 
The telescoping sum can be readily simplified and we can also write
\begin{equation}
 V_{m}=\sum_\ell P^{(+)}_\ell \Bigl(\prod_{j=1}^{my} P^{(-)}_{\ell+j}\Bigr)\Bigl[ 1-\Bigl(\prod_{j'=1}^{y} P^{(-)}_{\ell+my+j'}\Bigr)\Bigr]
\end{equation}
\subsection{Class II}
We now consider diagonal operators  that are still characterised by their local action around a particle, this time however  the number of spins down both after and before the particle are important. 
Few examples of their local action is reported in Table~\ref{t:3}. 
\begin{table*}[!htbp]
\centering
\begin{tabular}{c|c}
&$\ket{\dots (\uparrow)\underbrace{\downarrow\dots\downarrow}_{n_-}\uparrow_{j}\underbrace{\downarrow\dots\downarrow}_{n_+}(\uparrow)\dots}$\\
\hline
$ \sigma^z\otimes\frac{ 1+\sigma^z}{2}\otimes \sigma^z$&$(1-2\delta_{n_-,0})(1-2\delta_{n_+,0})$\\
$ \sigma^z\otimes \frac{ 1-\sigma^z}{2}\otimes \frac{ 1+\sigma^z}{2}\otimes  \sigma^z$&$(2\delta_{n_-,1}+\delta_{n_-,0}-1)(2\delta_{n_+,0}-1)$\\
$ \sigma^z\otimes \frac{ 1+\sigma^z}{2}\otimes \frac{ 1-\sigma^z}{2}\otimes  \sigma^z$&$(2\delta_{n_-,0}-1)(2\delta_{n_+,1}+\delta_{n_+,0}-1)$\\
$ \sigma^z\otimes \frac{ 1-\sigma^z}{2}\otimes \frac{ 1-\sigma^z}{2}\otimes \frac{ 1+\sigma^z}{2}\otimes  \sigma^z$&$(2\delta_{n_-,2}+\delta_{n_-,1}+\delta_{n_-,0}-1)(2\delta_{n_+,0}-1)$\\
$ \sigma^z\otimes \frac{ 1-\sigma^z}{2}\otimes \frac{ 1+\sigma^z}{2}\otimes \frac{ 1-\sigma^z}{2}\otimes  \sigma^z$&$(2\delta_{n_-,1}+\delta_{n_-,0}-1)(2\delta_{n_+,1}+\delta_{n_+,0}-1)$\\
$ \sigma^z\otimes \frac{ 1+\sigma^z}{2}\otimes \frac{ 1-\sigma^z}{2}\otimes \frac{ 1-\sigma^z}{2}\otimes  \sigma^z$&$(2\delta_{n_-,0}-1)(2\delta_{n_+,2}+\delta_{n_+,1}+\delta_{n_+,0}-1)$
\end{tabular}\caption{Local contribution to the eigenvalues of some class II operators, ordered by their range.}\label{t:3}
\end{table*}
The duality transformation can be worked out using that, in the pseudospace, \eqref{eq:classII} holds. 
\begin{figure*}[!htb]
\begin{multline}\label{eq:classII}
\frac{1- \tau_{j-k_1-1}^z}{2}\Bigl(\prod_{j'=1}^{k_1}\frac{1+ \tau_{j-j'}^z}{2}\Bigr)\frac{1- \tau_j^z}{2} \Bigl(\prod_{j''=1}^{k_2}\frac{1+ \tau_{j+j''}^z}{2}\Bigr) \frac{1- \tau_{j+k_2+1}^z}{2}\\
\ket{\dots(\downarrow^{b+n_-+1\!\!\!\pmod y})\underbrace{\uparrow\dots\uparrow}_{\lfloor \frac{n_-}{y}\rfloor} \downarrow_j^b \underbrace{\uparrow\dots\uparrow}_{\lfloor \frac{n_+}{y}\rfloor}(\downarrow^{b-n_+-1\!\!\!\pmod y})\dots}=\\
\delta_{\lfloor\frac{n_-}{y}\rfloor,k_1}\delta_{\lfloor\frac{n_+}{y}\rfloor,k_2}\ket{\dots(\downarrow^{b+n_-+1\!\!\!\pmod y})\underbrace{\uparrow\dots\uparrow}_{\lfloor \frac{n_-}{y}\rfloor} \downarrow_j^b \underbrace{\uparrow\dots\uparrow}_{\lfloor \frac{n_+}{y}\rfloor}(\downarrow^{b-n_+-1\!\!\!\pmod y})\dots}\, .
\end{multline}
\end{figure*}
Indeed, besides contributions that we worked out also for class I operators, class II operators are characterised by an additional one (cf. Table~\ref{t:3}):
\begin{description}
\item[Product of two Kronecker deltas $\delta_{n_-,k_1}\delta_{n_+,k_1}$] This kind of contributions can be worked out by properly multiplying the representation of the delta in \eqref{eq:deltan}:
\begin{equation}
\delta_{n_-,k_1}\delta_{n_+,k_2}=\delta_{\lfloor\frac{n_-}{y}\rfloor,\lfloor\frac{k_1}{y}\rfloor}\delta_{\lfloor\frac{n_+}{y}\rfloor, \lfloor\frac{k_2}{y}\rfloor}
\frac{1}{y^2}\sum_{m,m'=0}^{y-1} e^{2\pi i m\frac{b_{\tilde j}-b_{\tilde j+1}-1-k_2}{y}}e^{2\pi i m'\frac{b_{\tilde j-1}-b_{\tilde j}-1-k_1}{y}}
\end{equation}
Thus we find
\begin{equation}\label{eq:deltatoTT}
\delta_{n_-,k_1}\delta_{n_+,k_2}\rightarrow \sum_{j\in I}( Z^{(\lfloor\frac{k_1}{y},\lfloor\frac{k_2}{y}\rfloor)}_{1+k_1\!\!\!\!\!\pmod y,1+k_2\!\!\!\!\!\pmod y})_j\, ,
\end{equation}
where $( Z^{(n_1,n_2)}_{s_1\!\!\!\!\!\pmod y,s_2\!\!\!\!\!\pmod y})_j$ is the semilocal operator 
\begin{multline}\label{eq:TT}
( Z^{(n_1,n_2)}_{s_1\!\!\!\!\!\pmod y,s_2\!\!\!\!\!\pmod y})_j=
\int_{-\pi}^\pi\frac{\mathrm d \omega}{2\pi}\sum_{k\in B} \sum_{m_1,m_2=0}^{y-1} \frac{e^{2\pi i \frac{m_1(b_{k}-b_{k+1}-s_1)+m_2(b_{k+1}-b_{k+2}-s_2)}{y}}}{y^2}
 e^{i\omega (k+\sum_{\tilde j}^01-\sum_{j'}^j\frac{ 1-\tau^z_j}{2})}\\
\frac{ 1-\tau_j^z}{2}\Bigl(\prod_{m=1}^{n_1-1}\frac{ 1+\tau_{j+m}^z}{2}\Bigr)\frac{ 1-\tau_{j+n_1}^z}{2}\Bigl(\prod_{m=1}^{n_2-1}\frac{ 1+\tau_{j+n_1+m}^z}{2}\Bigr)\frac{ 1-\tau_{j+n_1+n_2}^z}{2}\, .
\end{multline}
\end{description}
We point out that identity~\eqref{eq:sumT} is  generalised in that the sum over all $s$ results in a (strictly) local operator.
\subsection{Class III and beyond}
The explicit calculations for the operators in classes I and II have revealed the general structure  of the mapping. Specifically, an operator with a density consisting of the tensor product of projectors $ P_\pm$ sandwiched between two $\sigma^z$ is class $M$ if there is a number $M-1$ of $ P_+$. The eigenvalue will then be written as a linear combination of Kronecker deltas and constants, the highest product involving $M$ Kronecker deltas. The species associated with the deltas appearing in the corresponding representation \eqref{eq:deltan} should be chosen according to which species surround the associated domains of spins down. The product of deltas is then represented by generalisations of  \eqref{eq:T} along the line of \eqref{eq:TT}. 

\section{Integrable representation in the presence of two impurities}\label{a:Bethe}
In this section we provide more details about the two-particle sector of the symmetric Hamiltonian shown in \eqref{eq:H}.
We have
\begin{multline}
:\tilde{ H}:\ket{n_1,n_2}=
 2J[\ket{n_1-1,n_2}+(1-\delta_{n_2,n_1+1})(\ket{n_1+1,n_2}\\
 +\ket{n_1,n_2-1})+\ket{n_1,n_2+1}]+4J\Delta(\delta_{n_2,n_1+1}-2)\ket{n_1,n_2}\\
 +2Jg\delta_{n_2,n_1+1}(\ket{n_1-1,n_1}+\ket{n_1+1,n_1+2})\, ,
\end{multline}
where the sequence $\underline{b}$ is understood. 
By inspecting the single-impurity problem, we identify the following single-particle eigenstates
\begin{equation}
\ket{\Psi(p)}=\sum_n e^{i n p}\ket{n}
\end{equation}
normalised according to $\braket{\Psi(p_1)|\Psi(p_2)}=2\pi\delta(p_1-p_2)$; the corresponding excitation energy reads
\begin{equation}
E(p)=4 J(\cos p-\Delta)\, .
\end{equation}
The Bethe Ansatz for the two-impurity problem can be written as follows
\begin{equation}
\ket{\Psi(p_1,p_2)}\sim \sum_{n_1<n_2} [e^{in_1 p_1+in_2 p_2}+S(p_1,p_2)e^{in_1 p_2+in_2 p_1}]\ket{n_1,n_2}
\end{equation}
where $S(p_1,p_2)$ is the scattering phase.
Enforcing the secular equation gives
\begin{equation}
S(p_1,p_2)=-\frac{1-2[\Delta +g\cos(p_1+p_2)]e^{i p_2}+e^{i (p_1+p_2)}}{1-2[\Delta+g\cos(p_1+p_2)]e^{i p_1}+e^{i (p_1+p_2)}}\, .
\end{equation}
We note that, for $g\neq 0$, it is not possible to define rapidities $\lambda_{1,2}$ so as to make the scattering phase of the difference form $S(\lambda_1-\lambda_2)$.
For the sake of transparency, we apply
an artificial finite-volume regularisation with periodic boundary conditions. The momenta satisfy in turn the Bethe equations
\begin{equation}
e^{iLp_1}=S(p_2,p_1)\qquad
e^{iLp_2}=S(p_1,p_2)
\end{equation}
The first effect of interaction is that these equations exhibit both real and complex solutions, the latter ones describing bound states of two particles. Specifically, the real solutions differ from the noninteracting ones in a negligible $O(1/L)$ correction that is nevertheless responsible for the incompleteness of the spanned space (it leaves out a one-dimensional space of solutions in the two-dimensional space of momenta). 

In the following we evaluate the late time behaviour of the two cumulative distributions
\begin{equation}\label{eq:cumdist21}
F_t^{(1)}(n)=\sum_{n_1\leq n }p_{n_1,t}^{(1)}=\sum_{n_1\leq n}\sum_{n_2>n_1}|\braket{n_1,n_2|e^{-i\tilde{ H}_{eff} t }\tilde{ P}|\Psi_0}|^2\, .
\end{equation}
\begin{equation}\label{eq:cumdist22}
F_t^{(2)}(n)=\sum_{n_2\leq n }p_{n_2,t}^{(2)}=\sum_{n_1<n_2\leq n}|\braket{n_1,n_2|e^{-i\tilde{ H}_{eff} t }\tilde{ P}|\Psi_0}|^2\, .
\end{equation} 

\subsection{Real solutions}
We start by considering the contribution from the real solutions.
Let us define the non-normalised states in a symmetric way 
\begin{equation}
\ket{\Psi(p_1,p_2)}=
\sum_{-\frac{L}{2}< n_1<n_2\leq \frac{L}{2}} \!\!\!\!\!\!\!\!\![A_{p_1,p_2}e^{in_1 p_1+in_2 p_2}-A_{p_2,p_1}e^{in_1 p_2+in_2 p_1}]\ket{n_1,n_2}
\end{equation}
where
\begin{equation}
A_{p_1,p_2}=1-2[\Delta+g\cos(p_1+p_2)]e^{i p_1}+e^{i (p_1+p_2)}\, .
\end{equation}
\begin{equation}
(A_{-p_1,-p_2}=A_{p_2,p_1}e^{-i(p_1+p_2)})
\end{equation}
In the thermodynamic limit such states are normalised according to 
\begin{equation}
\braket{\Psi(p_1,p_2)|\Psi(p_1',p_2')}=(2\pi)^2\delta(p_1'-p_1)\delta(p_2'-p_2) |A(p_1,p_2)|^2\, ,
\end{equation}
with finite-size corrections that scale as $L^{-1}$. In addition, their density is asymptotically flat, indeed, for a smooth test function $f(p_j)$ and $\epsilon$ approaching zero more slowly than $1/L$, we have ($\bar j=3-j$)
\begin{equation}
\sum_{p_j\atop
\mathrm{Bethe\ eqs}}f(p_j)=\oint_{[-\pi,\pi]_\epsilon}\frac{\mathrm d p}{2\pi }\frac{L-i\partial_{p}\log S(p,p_{\bar j})}{1-e^{-iL p}S(p_{\bar j},p)}f(p)
\rightarrow L\int_{-\pi}^\pi\frac{\mathrm d p}{2\pi}[1-\frac{i\partial_p\log S(p,p_{\bar j})}{L}]f(p)\, ,
\end{equation}
where $[-\pi,\pi]_\epsilon$ is the rectangular contour surrounding the interval $[-\pi,\pi]$ with imaginary  part $\pm i\epsilon$. 
We can conclude that the density is flat observing that the correction associated with the scattering phase approaches zero as $O(L^{-1})$. We are now in a position to compute the contribution $F_{t,\mathrm{m}}^{(j)}(n)$ to $F_{t,\mathrm{m}}^{(j)}(n)$ of the excitations with real momenta.
The antisymmetry of the state under the interchange of $p_1$ and $p_2$ allows us to express them as follows
\begin{multline}
F_{t, \mathrm{m}}^{(1)}(n)= \frac{1}{4}\sum_{m_1= 0}^\infty\sum_{m=1}^\infty
\Bigl|\int\frac{\mathrm d^2 p}{(2\pi)^2}e^{-i[E(p_1)+E(p_2)]t}
\frac{\braket{-m_1+n,-m_1+n+m|\Psi(p_1,p_2)}\braket{\Psi(p_1,p_2)|n_1^{(0)},n_2^{(0)}}}{|A(p_1,p_2)|^2}\Bigr|^2= \\
 \frac{1}{2}\int_{-\pi}^\pi\frac{\mathrm d^2 p\mathrm d ^2p'}{(2\pi)^4}\frac{e^{-i[E(p_1)+E(p_2)-E(p_1')-E(p_2')]t}e^{in(p_1+p_2-p_1'-p_2')}}{1-e^{-i(p_1+p_2-p_1'-p_2'-i\epsilon)}}
\frac{\braket{\Psi(p_1,p_2)|n_1^{(0)},n_2^{(0)}}\braket{n_1^{(0)},n_2^{(0)}|\Psi(p_1',p_2')}}{|A(p_1,p_2)|^2|A(p_1',p_2')|^2}\\
\times\Bigl[\frac{A(p_1,p_2)A(-p_1',-p_2')}{e^{-i (p_2-p_2'+i\epsilon)}-1}-\frac{A(p_2,p_1)A(-p_1',-p_2')}{e^{-i(p_1-p_2'+i\epsilon)}-1}\Bigr]
\end{multline}
It is convenient to shift the integration variables in such a way that only one momentum appears in the phases at the denominators. This, in turn, allows us to carry them out separately; specifically we get
\begin{multline}
F_{t, \mathrm{m}}^{(1)}(n)= \frac{1}{2}\int_{-\pi}^\pi\frac{\mathrm d^2 p\mathrm d ^2p'}{(2\pi)^4}\frac{e^{-i[E(p_1)+E(p_2)-E(p_1'+p_1-p_2')-E(p_2'+p_2)]t}e^{-in p_1'}}{1-e^{i(p_1'+i\epsilon)}}\\
\frac{\braket{\Psi(p_1,p_2)|n_1^{(0)},n_2^{(0)}}\braket{n_1^{(0)},n_2^{(0)}|\Psi(p_1'+p_1-p_2',p_2'+p_2)}}{|A(p_1,p_2)|^2|A(p_1'+p_1-p_2',p_2'+p_2)|^2}\frac{A(p_1,p_2)A(-p_1'-p_1+p_2',-p_2'-p_2)}{e^{i (p_2'-i\epsilon)}-1}\\
-\frac{1}{2}\int_{-\pi}^\pi\frac{\mathrm d^2 p\mathrm d ^2p'}{(2\pi)^4}\frac{e^{-i[E(p_1)+E(p_2)-E(p_1'+p_2-p_2')-E(p_2'+p_1)]t}e^{-in p_1'}}{1-e^{i(p_1'+i\epsilon)}}\\
\frac{\braket{\Psi(p_1,p_2)|n_1^{(0)},n_2^{(0)}}\braket{n_1^{(0)},n_2^{(0)}|\Psi(p_1'+p_2-p_2',p_2'+p_1)}}{|A(p_1,p_2)|^2|A(p_1'+p_2-p_2',p_2'+p_1)|^2}\frac{A(p_2,p_1)A(-p_1'-p_2+p_2',-p_2'-p_1)}{e^{i(p_2'-i\epsilon)}-1}\\
\sim \frac{1}{2}\int_{-\pi}^\pi\frac{\mathrm d^2 p}{(2\pi)^2}\theta_H(n-\min(v(p_1),v(p_2))t)\frac{\braket{\Psi(p_1,p_2)|n_1^{(0)},n_2^{(0)}}\braket{n_1^{(0)},n_2^{(0)}|\Psi(p_1,p_2)}}{|A(p_1,p_2)|^2}
\end{multline}

Thus we have
\begin{equation}
F_{t, \mathrm{m}}^{(1)}(n)\sim \frac{1}{2}\int_{-\pi}^\pi\frac{\mathrm d^2 p}{(2\pi)^2}\theta_H(n-\min(v(p_1),v(p_2))t)|e^{i(n_1^{(0)} -n_2^{(0)})(p_1-p_2)}+S(p_1,p_2)|^2
\end{equation}
and analogously for $F_{t, \mathrm{m}}^{(2)}(n)$
\begin{equation}
F_{t, \mathrm{m}}^{(2)}(n)\sim \frac{1}{2}\int_{-\pi}^\pi\frac{\mathrm d^2 p}{(2\pi)^2}\theta_H(n-\max(v(p_1),v(p_2))t)|e^{i(n_1^{(0)} -n_2^{(0)})(p_1-p_2)}+S(p_1,p_2)|^2\, .
\end{equation}
The incompleteness of the real solutions is evident if we take the limit $n\rightarrow\infty$, which gives
\begin{equation}\label{eq:displacement}
\lim_{n\rightarrow\infty}F_{t, \mathrm{m}}^{(j)}(n)\sim 1+\mathrm{Re}\int_{-\pi}^\pi\frac{\mathrm d^2 p}{(2\pi)^2}e^{i(n_2^{(0)} -n_1^{(0)})(p_1-p_2)}S(p_1,p_2)
\end{equation}
We will see before long that the bound state contribution cancels out the displacement from $1$.

\subsection{Bound states}
Except for the region $g=\Delta\in(-\frac{1}{2},\frac{1}{2})$, the Bethe equations exhibit always complex solutions. In the simple case we are considering, such solutions can be easily worked out imposing that, in the Bethe equations , $|e^{iLp_j}|$  approaches either $0$ or $\infty$. 
In particular, we find that 
the bound state can have any momentum 
within the region
\begin{equation}
1+\cos p<2(\Delta + g \cos p)^2\, ,
\end{equation}
i.e. 
\begin{equation}
\cos p>\frac{1-4 g\Delta+\sqrt{1+8g^2-8 g\Delta}}{4 g^2}\quad \vee\quad  \cos p<\frac{1-4 g\Delta-\sqrt{1+8g^2-8 g\Delta}}{4 g^2}\, .
\end{equation}
Note that this condition is always satisfied when $g\Delta>\frac{1}{8}+ g^2$ or $1<(\Delta+g)^2\wedge 4g(\Delta+g)<1$. 
The corresponding bound state reads
\begin{equation}
\ket{\Psi_2(p)}\sim \sqrt{\Bigl(\frac{\Delta +g \cos  p}{\cos \frac{p}{2}}\Big)^2-1}\sum_{n_1<n_2}e^{i\frac{n_1+n_2}{2}p} \Bigl(\frac{\cos\frac{ p}{2}}{ \Delta +g  \cos p}\Bigr)^{n_2-n_1}\ket{n_1,n_2}\, ,
\end{equation}
where we enforced the standard normalisation  $\braket{\Psi( p')|\Psi( p)}=2\pi\delta( p- p')$. Its energy is given by
\begin{equation}
E_2(p)=4g\cos p+4J[\frac{\cos^2\frac{ p}{2}}{\Delta +g\cos  p}-\Delta]
\end{equation}
We can now compute the bound-state contribution to the cumulative distributions. We have
\begin{multline}
F_{t,\mathrm{b}}^{(1)}(n)=\\
\sum_{m_1= 0}^\infty\sum_{m=1}^\infty|\int_{-\pi}^\pi\frac{\mathrm d p}{2\pi}\theta_H(1+\cos p<2(\Delta + g \cos p)^2)e^{-i E_2(p)t}
\braket{-m_1+n,-m_1+n+m|\Psi_2(p)}\braket{\Psi_2(p)|n_1^{(0)},n_2^{(0)}}|^2=\\
 \int_{-\pi}^\pi\frac{\mathrm d p\mathrm d p'}{(2\pi)^2}\braket{\Psi_2(p)|n_1^{(0)},n_2^{(0)}}
\braket{n_1^{(0)},n_2^{(0)}|\Psi_2(p'+p)}\theta_H(\cos^2\tfrac{p}{2}<(\Delta +g \cos p)^2)\theta_H(\cos^2 \tfrac{p'+p}{2}<(\Delta + g \cos (p'+p))^2)\\
\sqrt{\Bigl(\tfrac{\Delta +g \cos  p}{\cos \frac{p}{2}}\Big)^2-1}\sqrt{\Bigl(\tfrac{\Delta +g \cos  (p'+p)}{\cos \frac{p'+p}{2}}\Big)^2-1}\frac{e^{-i n p'}}{1-e^{i(p'+i\epsilon)}} \frac{e^{-i [E_2(p)-E_2(p'+p)]t}}{\frac{[\Delta +g  \cos p][\Delta +g  \cos (p'+p)]}{e^{-i\frac{p'}{2}}\cos\frac{ p}{2}\cos\frac{ p'+p}{2}}-1}
\sim\\
\int_{-\pi}^\pi\frac{\mathrm d p}{2\pi}\theta_H(1+\cos p<2(\Delta + g\cos p)^2)\theta_H(n-v_2(p))\braket{\Psi_2(p)|n_1^{(0)},n_2^{(0)}}
\braket{n_1^{(0)},n_2^{(0)}|\Psi_2(p)}
\end{multline}
Thus we have
\begin{equation}
F_{t,\mathrm{b}}^{(1)}(n)\sim \int_{-\pi}^\pi\frac{\mathrm d p}{2\pi}\theta_H(n-v_2(p)t)\max\left[\Bigl(\frac{\Delta +g \cos  p}{\cos \frac{p}{2}}\Big)^2-1,0\right] \Bigl(\frac{\cos^2\frac{ p}{2}}{[\Delta +g  \cos p]^2}\Bigr)^{n_2^{(0)}-n_1^{(0)}}\, .
\end{equation}
Concerning $F_{t,\mathrm{b}}^{(2)}(n)$, as one could expect from physical arguments based on the fact that the impurities are bounded, we find $F_{t,\mathrm{b}}^{(2)}(t)\sim F_{t,\mathrm{b}}^{(1)}(n)$, which we can therefore simply call $F_{t,\mathrm{b}}(n)$.

Importantly, despite the different appearance of the integrals, the limit $n\rightarrow\infty$ exactly compensates the displacement from $1$ that we got in \eqref{eq:displacement} considering only the real solutions, supporting the completeness of the Bethe states identified in this sector.

\twocolumngrid

\bibliography{references}

\begin{thebibliography}{57}%
\makeatletter
\providecommand \@ifxundefined [1]{%
 \@ifx{#1\undefined}
}%
\providecommand \@ifnum [1]{%
 \ifnum #1\expandafter \@firstoftwo
 \else \expandafter \@secondoftwo
 \fi
}%
\providecommand \@ifx [1]{%
 \ifx #1\expandafter \@firstoftwo
 \else \expandafter \@secondoftwo
 \fi
}%
\providecommand \natexlab [1]{#1}%
\providecommand \enquote  [1]{``#1''}%
\providecommand \bibnamefont  [1]{#1}%
\providecommand \bibfnamefont [1]{#1}%
\providecommand \citenamefont [1]{#1}%
\providecommand \href@noop [0]{\@secondoftwo}%
\providecommand \href [0]{\begingroup \@sanitize@url \@href}%
\providecommand \@href[1]{\@@startlink{#1}\@@href}%
\providecommand \@@href[1]{\endgroup#1\@@endlink}%
\providecommand \@sanitize@url [0]{\catcode `\\12\catcode `\$12\catcode
  `\&12\catcode `\#12\catcode `\^12\catcode `\_12\catcode `\%12\relax}%
\providecommand \@@startlink[1]{}%
\providecommand \@@endlink[0]{}%
\providecommand \url  [0]{\begingroup\@sanitize@url \@url }%
\providecommand \@url [1]{\endgroup\@href {#1}{\urlprefix }}%
\providecommand \urlprefix  [0]{URL }%
\providecommand \Eprint [0]{\href }%
\providecommand \doibase [0]{http://dx.doi.org/}%
\providecommand \selectlanguage [0]{\@gobble}%
\providecommand \bibinfo  [0]{\@secondoftwo}%
\providecommand \bibfield  [0]{\@secondoftwo}%
\providecommand \translation [1]{[#1]}%
\providecommand \BibitemOpen [0]{}%
\providecommand \bibitemStop [0]{}%
\providecommand \bibitemNoStop [0]{.\EOS\space}%
\providecommand \EOS [0]{\spacefactor3000\relax}%
\providecommand \BibitemShut  [1]{\csname bibitem#1\endcsname}%
\let\auto@bib@innerbib\@empty
\bibitem [{\citenamefont {Giamarchi}(2003)}]{Giamarchi2003Quantum}%
  \BibitemOpen
  \bibfield  {author} {\bibinfo {author} {\bibfnamefont {T.}~\bibnamefont
  {Giamarchi}},\ }\href {\doibase 10.1093/acprof:oso/9780198525004.001.0001}
  {\emph {\bibinfo {title} {{{Quantum Physics in One Dimension}}}}}\ (\bibinfo
  {publisher} {Oxford University Press},\ \bibinfo {year} {2003})\BibitemShut
  {NoStop}%
\bibitem [{\citenamefont {Essler}\ \emph {et~al.}(2005)\citenamefont {Essler},
  \citenamefont {Frahm}, \citenamefont {Göhmann}, \citenamefont {Klümper},\
  and\ \citenamefont {Korepin}}]{Essler2005The}%
  \BibitemOpen
  \bibfield  {author} {\bibinfo {author} {\bibfnamefont {F.~H.~L.}\
  \bibnamefont {Essler}}, \bibinfo {author} {\bibfnamefont {H.}~\bibnamefont
  {Frahm}}, \bibinfo {author} {\bibfnamefont {F.}~\bibnamefont {Göhmann}},
  \bibinfo {author} {\bibfnamefont {A.}~\bibnamefont {Klümper}}, \ and\
  \bibinfo {author} {\bibfnamefont {V.~E.}\ \bibnamefont {Korepin}},\ }\href
  {\doibase 10.1017/CBO9780511534843} {\emph {\bibinfo {title} {{The
  One-Dimensional Hubbard Model}}}}\ (\bibinfo  {publisher} {Cambridge
  University Press},\ \bibinfo {year} {2005})\BibitemShut {NoStop}%
\bibitem [{\citenamefont {Zeng}\ \emph {et~al.}(2019)\citenamefont {Zeng},
  \citenamefont {Chen}, \citenamefont {Zhou},\ and\ \citenamefont
  {Wen}}]{Zeng2019Quantum}%
  \BibitemOpen
  \bibfield  {author} {\bibinfo {author} {\bibfnamefont {B.}~\bibnamefont
  {Zeng}}, \bibinfo {author} {\bibfnamefont {X.}~\bibnamefont {Chen}}, \bibinfo
  {author} {\bibfnamefont {D.-L.}\ \bibnamefont {Zhou}}, \ and\ \bibinfo
  {author} {\bibfnamefont {X.-G.}\ \bibnamefont {Wen}},\ }\href {\doibase
  10.1007/978-1-4939-9084-9} {\emph {\bibinfo {title} {{{Quantum Information
  Meets Quantum Matter}}}}}\ (\bibinfo  {publisher} {Oxford University Press},\
  \bibinfo {year} {2019})\BibitemShut {NoStop}%
\bibitem [{\citenamefont {Bloch}\ \emph {et~al.}(2008)\citenamefont {Bloch},
  \citenamefont {Dalibard},\ and\ \citenamefont
  {Zwerger}}]{Bloch2008Many-body}%
  \BibitemOpen
  \bibfield  {author} {\bibinfo {author} {\bibfnamefont {I.}~\bibnamefont
  {Bloch}}, \bibinfo {author} {\bibfnamefont {J.}~\bibnamefont {Dalibard}}, \
  and\ \bibinfo {author} {\bibfnamefont {W.}~\bibnamefont {Zwerger}},\
  }\bibfield  {title} {\enquote {\bibinfo {title} {{Many-body physics with
  ultracold gases}},}\ }\href {\doibase 10.1103/RevModPhys.80.885} {\bibfield
  {journal} {\bibinfo  {journal} {Rev. Mod. Phys.}\ }\textbf {\bibinfo {volume}
  {80}},\ \bibinfo {pages} {885} (\bibinfo {year} {2008})}\BibitemShut
  {NoStop}%
\bibitem [{\citenamefont {Cazalilla}\ \emph {et~al.}(2011)\citenamefont
  {Cazalilla}, \citenamefont {Citro}, \citenamefont {Giamarchi}, \citenamefont
  {Orignac},\ and\ \citenamefont {Rigol}}]{Cazalilla2011One}%
  \BibitemOpen
  \bibfield  {author} {\bibinfo {author} {\bibfnamefont {M.~A.}\ \bibnamefont
  {Cazalilla}}, \bibinfo {author} {\bibfnamefont {R.}~\bibnamefont {Citro}},
  \bibinfo {author} {\bibfnamefont {T.}~\bibnamefont {Giamarchi}}, \bibinfo
  {author} {\bibfnamefont {E.}~\bibnamefont {Orignac}}, \ and\ \bibinfo
  {author} {\bibfnamefont {M.}~\bibnamefont {Rigol}},\ }\bibfield  {title}
  {\enquote {\bibinfo {title} {{One dimensional bosons: From condensed matter
  systems to ultracold gases}},}\ }\href {\doibase 10.1103/RevModPhys.83.1405}
  {\bibfield  {journal} {\bibinfo  {journal} {Rev. Mod. Phys.}\ }\textbf
  {\bibinfo {volume} {83}},\ \bibinfo {pages} {1405} (\bibinfo {year}
  {2011})}\BibitemShut {NoStop}%
\bibitem [{\citenamefont {Buluta}\ and\ \citenamefont
  {Nori}(2009)}]{Buluta2009Science}%
  \BibitemOpen
  \bibfield  {author} {\bibinfo {author} {\bibfnamefont {I.}~\bibnamefont
  {Buluta}}\ and\ \bibinfo {author} {\bibfnamefont {F.}~\bibnamefont {Nori}},\
  }\bibfield  {title} {\enquote {\bibinfo {title} {{Quantum Simulators}},}\
  }\href {\doibase 10.1126/science.1177838} {\bibfield  {journal} {\bibinfo
  {journal} {Science}\ }\textbf {\bibinfo {volume} {326}},\ \bibinfo {pages}
  {108} (\bibinfo {year} {2009})}\BibitemShut {NoStop}%
\bibitem [{\citenamefont {Cirac}\ and\ \citenamefont
  {Zoller}(2012)}]{Cirac2012Goals}%
  \BibitemOpen
  \bibfield  {author} {\bibinfo {author} {\bibfnamefont {J.~I.}\ \bibnamefont
  {Cirac}}\ and\ \bibinfo {author} {\bibfnamefont {P.}~\bibnamefont {Zoller}},\
  }\bibfield  {title} {\enquote {\bibinfo {title} {{Goals and opportunities in
  quantum simulation}},}\ }\href {\doibase 10.1038/nphys2275} {\bibfield
  {journal} {\bibinfo  {journal} {Nature Physics}\ }\textbf {\bibinfo {volume}
  {8}},\ \bibinfo {pages} {264} (\bibinfo {year} {2012})}\BibitemShut {NoStop}%
\bibitem [{\citenamefont {Bernien}\ \emph {et~al.}(2017)\citenamefont
  {Bernien}, \citenamefont {Schwartz}, \citenamefont {Keesling}, \citenamefont
  {Levine}, \citenamefont {Omran}, \citenamefont {Pichler}, \citenamefont
  {Choi}, \citenamefont {Zibrov}, \citenamefont {Endres}, \citenamefont
  {Greiner}, \citenamefont {Vuleti{\'{c}}},\ and\ \citenamefont
  {Lukin}}]{Bernien2017Probing}%
  \BibitemOpen
  \bibfield  {author} {\bibinfo {author} {\bibfnamefont {H.}~\bibnamefont
  {Bernien}}, \bibinfo {author} {\bibfnamefont {S.}~\bibnamefont {Schwartz}},
  \bibinfo {author} {\bibfnamefont {A.}~\bibnamefont {Keesling}}, \bibinfo
  {author} {\bibfnamefont {H.}~\bibnamefont {Levine}}, \bibinfo {author}
  {\bibfnamefont {A.}~\bibnamefont {Omran}}, \bibinfo {author} {\bibfnamefont
  {H.}~\bibnamefont {Pichler}}, \bibinfo {author} {\bibfnamefont
  {S.}~\bibnamefont {Choi}}, \bibinfo {author} {\bibfnamefont {A.~S.}\
  \bibnamefont {Zibrov}}, \bibinfo {author} {\bibfnamefont {M.}~\bibnamefont
  {Endres}}, \bibinfo {author} {\bibfnamefont {M.}~\bibnamefont {Greiner}},
  \bibinfo {author} {\bibfnamefont {V.}~\bibnamefont {Vuleti{\'{c}}}}, \ and\
  \bibinfo {author} {\bibfnamefont {M.~D.}\ \bibnamefont {Lukin}},\ }\bibfield
  {title} {\enquote {\bibinfo {title} {{Probing many-body dynamics on a 51-atom
  quantum simulator}},}\ }\href {\doibase 10.1038/nature24622} {\bibfield
  {journal} {\bibinfo  {journal} {Nature}\ }\textbf {\bibinfo {volume} {551}},\
  \bibinfo {pages} {579} (\bibinfo {year} {2017})}\BibitemShut {NoStop}%
\bibitem [{\citenamefont {Polkovnikov}\ \emph {et~al.}(2011)\citenamefont
  {Polkovnikov}, \citenamefont {Sengupta}, \citenamefont {Silva},\ and\
  \citenamefont {Vengalattore}}]{Polkovnikov2011Colloquium}%
  \BibitemOpen
  \bibfield  {author} {\bibinfo {author} {\bibfnamefont {A.}~\bibnamefont
  {Polkovnikov}}, \bibinfo {author} {\bibfnamefont {K.}~\bibnamefont
  {Sengupta}}, \bibinfo {author} {\bibfnamefont {A.}~\bibnamefont {Silva}}, \
  and\ \bibinfo {author} {\bibfnamefont {M.}~\bibnamefont {Vengalattore}},\
  }\bibfield  {title} {\enquote {\bibinfo {title} {{Colloquium: Nonequilibrium
  dynamics of closed interacting quantum systems}},}\ }\href {\doibase
  10.1103/RevModPhys.83.863} {\bibfield  {journal} {\bibinfo  {journal} {Rev.
  Mod. Phys.}\ }\textbf {\bibinfo {volume} {83}},\ \bibinfo {pages} {863}
  (\bibinfo {year} {2011})}\BibitemShut {NoStop}%
\bibitem [{\citenamefont {Gogolin}\ and\ \citenamefont
  {Eisert}(2016)}]{Gogolin2016Equilibration}%
  \BibitemOpen
  \bibfield  {author} {\bibinfo {author} {\bibfnamefont {C.}~\bibnamefont
  {Gogolin}}\ and\ \bibinfo {author} {\bibfnamefont {J.}~\bibnamefont
  {Eisert}},\ }\bibfield  {title} {\enquote {\bibinfo {title} {{Equilibration,
  thermalisation, and the emergence of statistical mechanics in closed quantum
  systems}},}\ }\href {\doibase 10.1088/0034-4885/79/5/056001} {\bibfield
  {journal} {\bibinfo  {journal} {Reports on Progress in Physics}\ }\textbf
  {\bibinfo {volume} {79}},\ \bibinfo {pages} {056001} (\bibinfo {year}
  {2016})}\BibitemShut {NoStop}%
\bibitem [{\citenamefont {Gring}\ \emph {et~al.}(2012)\citenamefont {Gring},
  \citenamefont {Kuhnert}, \citenamefont {Langen}, \citenamefont {Kitagawa},
  \citenamefont {Rauer}, \citenamefont {Schreitl}, \citenamefont {Mazets},
  \citenamefont {Smith}, \citenamefont {Demler},\ and\ \citenamefont
  {Schmiedmayer}}]{Gring2012Relaxation}%
  \BibitemOpen
  \bibfield  {author} {\bibinfo {author} {\bibfnamefont {M.}~\bibnamefont
  {Gring}}, \bibinfo {author} {\bibfnamefont {M.}~\bibnamefont {Kuhnert}},
  \bibinfo {author} {\bibfnamefont {T.}~\bibnamefont {Langen}}, \bibinfo
  {author} {\bibfnamefont {T.}~\bibnamefont {Kitagawa}}, \bibinfo {author}
  {\bibfnamefont {B.}~\bibnamefont {Rauer}}, \bibinfo {author} {\bibfnamefont
  {M.}~\bibnamefont {Schreitl}}, \bibinfo {author} {\bibfnamefont
  {I.}~\bibnamefont {Mazets}}, \bibinfo {author} {\bibfnamefont {D.~A.}\
  \bibnamefont {Smith}}, \bibinfo {author} {\bibfnamefont {E.}~\bibnamefont
  {Demler}}, \ and\ \bibinfo {author} {\bibfnamefont {J.}~\bibnamefont
  {Schmiedmayer}},\ }\bibfield  {title} {\enquote {\bibinfo {title}
  {{Relaxation and Prethermalization in an Isolated Quantum System}},}\ }\href
  {\doibase 10.1126/science.1224953} {\bibfield  {journal} {\bibinfo  {journal}
  {Science}\ }\textbf {\bibinfo {volume} {337}},\ \bibinfo {pages} {1318}
  (\bibinfo {year} {2012})}\BibitemShut {NoStop}%
\bibitem [{\citenamefont {Schemmer}\ \emph {et~al.}(2019)\citenamefont
  {Schemmer}, \citenamefont {Bouchoule}, \citenamefont {Doyon},\ and\
  \citenamefont {Dubail}}]{Schemmer2019Generalized}%
  \BibitemOpen
  \bibfield  {author} {\bibinfo {author} {\bibfnamefont {M.}~\bibnamefont
  {Schemmer}}, \bibinfo {author} {\bibfnamefont {I.}~\bibnamefont {Bouchoule}},
  \bibinfo {author} {\bibfnamefont {B.}~\bibnamefont {Doyon}}, \ and\ \bibinfo
  {author} {\bibfnamefont {J.}~\bibnamefont {Dubail}},\ }\bibfield  {title}
  {\enquote {\bibinfo {title} {{Generalized Hydrodynamics on an Atom Chip}},}\
  }\href {\doibase 10.1103/PhysRevLett.122.090601} {\bibfield  {journal}
  {\bibinfo  {journal} {Phys. Rev. Lett.}\ }\textbf {\bibinfo {volume} {122}},\
  \bibinfo {pages} {090601} (\bibinfo {year} {2019})}\BibitemShut {NoStop}%
\bibitem [{\citenamefont {Rigol}\ \emph {et~al.}(2007)\citenamefont {Rigol},
  \citenamefont {Dunjko}, \citenamefont {Yurovsky},\ and\ \citenamefont
  {Olshanii}}]{Rigol2007Relaxation}%
  \BibitemOpen
  \bibfield  {author} {\bibinfo {author} {\bibfnamefont {M.}~\bibnamefont
  {Rigol}}, \bibinfo {author} {\bibfnamefont {V.}~\bibnamefont {Dunjko}},
  \bibinfo {author} {\bibfnamefont {V.}~\bibnamefont {Yurovsky}}, \ and\
  \bibinfo {author} {\bibfnamefont {M.}~\bibnamefont {Olshanii}},\ }\bibfield
  {title} {\enquote {\bibinfo {title} {{Relaxation in a Completely Integrable
  Many-Body Quantum System: An Ab Initio Study of the Dynamics of the Highly
  Excited States of 1D Lattice Hard-Core Bosons}},}\ }\href {\doibase
  10.1103/PhysRevLett.98.050405} {\bibfield  {journal} {\bibinfo  {journal}
  {Phys. Rev. Lett.}\ }\textbf {\bibinfo {volume} {98}},\ \bibinfo {pages}
  {050405} (\bibinfo {year} {2007})}\BibitemShut {NoStop}%
\bibitem [{\citenamefont {Essler}\ and\ \citenamefont
  {Fagotti}(2016)}]{Essler2016Quench}%
  \BibitemOpen
  \bibfield  {author} {\bibinfo {author} {\bibfnamefont {F.~H.~L.}\
  \bibnamefont {Essler}}\ and\ \bibinfo {author} {\bibfnamefont
  {M.}~\bibnamefont {Fagotti}},\ }\bibfield  {title} {\enquote {\bibinfo
  {title} {{Quench dynamics and relaxation in isolated integrable quantum spin
  chains}},}\ }\href {\doibase 10.1088/1742-5468/2016/06/064002} {\bibfield
  {journal} {\bibinfo  {journal} {Journal of Statistical Mechanics: Theory and
  Experiment}\ }\textbf {\bibinfo {volume} {2016}},\ \bibinfo {pages} {064002}
  (\bibinfo {year} {2016})}\BibitemShut {NoStop}%
\bibitem [{\citenamefont {Ilievski}\ \emph {et~al.}(2016)\citenamefont
  {Ilievski}, \citenamefont {Medenjak}, \citenamefont {Prosen},\ and\
  \citenamefont {Zadnik}}]{Ilievski2016Quasilocal}%
  \BibitemOpen
  \bibfield  {author} {\bibinfo {author} {\bibfnamefont {E.}~\bibnamefont
  {Ilievski}}, \bibinfo {author} {\bibfnamefont {M.}~\bibnamefont {Medenjak}},
  \bibinfo {author} {\bibfnamefont {T.}~\bibnamefont {Prosen}}, \ and\ \bibinfo
  {author} {\bibfnamefont {L.}~\bibnamefont {Zadnik}},\ }\bibfield  {title}
  {\enquote {\bibinfo {title} {{Quasilocal charges in integrable lattice
  systems}},}\ }\href {\doibase 10.1088/1742-5468/2016/06/064008} {\bibfield
  {journal} {\bibinfo  {journal} {Journal of Statistical Mechanics: Theory and
  Experiment}\ }\textbf {\bibinfo {volume} {2016}},\ \bibinfo {pages} {064008}
  (\bibinfo {year} {2016})}\BibitemShut {NoStop}%
\bibitem [{\citenamefont {Ilievski}\ \emph {et~al.}(2015)\citenamefont
  {Ilievski}, \citenamefont {De~Nardis}, \citenamefont {Wouters}, \citenamefont
  {Caux}, \citenamefont {Essler},\ and\ \citenamefont
  {Prosen}}]{Ilievski2015Complete}%
  \BibitemOpen
  \bibfield  {author} {\bibinfo {author} {\bibfnamefont {E.}~\bibnamefont
  {Ilievski}}, \bibinfo {author} {\bibfnamefont {J.}~\bibnamefont {De~Nardis}},
  \bibinfo {author} {\bibfnamefont {B.}~\bibnamefont {Wouters}}, \bibinfo
  {author} {\bibfnamefont {J.-S.}\ \bibnamefont {Caux}}, \bibinfo {author}
  {\bibfnamefont {F.~H.~L.}\ \bibnamefont {Essler}}, \ and\ \bibinfo {author}
  {\bibfnamefont {T.}~\bibnamefont {Prosen}},\ }\bibfield  {title} {\enquote
  {\bibinfo {title} {{Complete Generalized Gibbs Ensembles in an Interacting
  Theory}},}\ }\href {\doibase 10.1103/PhysRevLett.115.157201} {\bibfield
  {journal} {\bibinfo  {journal} {Phys. Rev. Lett.}\ }\textbf {\bibinfo
  {volume} {115}},\ \bibinfo {pages} {157201} (\bibinfo {year}
  {2015})}\BibitemShut {NoStop}%
\bibitem [{\citenamefont {Castro-Alvaredo}\ \emph {et~al.}(2016)\citenamefont
  {Castro-Alvaredo}, \citenamefont {Doyon},\ and\ \citenamefont
  {Yoshimura}}]{Castro-Alvaredo2016Emergent}%
  \BibitemOpen
  \bibfield  {author} {\bibinfo {author} {\bibfnamefont {O.~A.}\ \bibnamefont
  {Castro-Alvaredo}}, \bibinfo {author} {\bibfnamefont {B.}~\bibnamefont
  {Doyon}}, \ and\ \bibinfo {author} {\bibfnamefont {T.}~\bibnamefont
  {Yoshimura}},\ }\bibfield  {title} {\enquote {\bibinfo {title} {{Emergent
  Hydrodynamics in Integrable Quantum Systems Out of Equilibrium}},}\ }\href
  {\doibase 10.1103/PhysRevX.6.041065} {\bibfield  {journal} {\bibinfo
  {journal} {Phys. Rev. X}\ }\textbf {\bibinfo {volume} {6}},\ \bibinfo {pages}
  {041065} (\bibinfo {year} {2016})}\BibitemShut {NoStop}%
\bibitem [{\citenamefont {Bertini}\ \emph {et~al.}(2016)\citenamefont
  {Bertini}, \citenamefont {Collura}, \citenamefont {De~Nardis},\ and\
  \citenamefont {Fagotti}}]{Bertini2016Transport}%
  \BibitemOpen
  \bibfield  {author} {\bibinfo {author} {\bibfnamefont {B.}~\bibnamefont
  {Bertini}}, \bibinfo {author} {\bibfnamefont {M.}~\bibnamefont {Collura}},
  \bibinfo {author} {\bibfnamefont {J.}~\bibnamefont {De~Nardis}}, \ and\
  \bibinfo {author} {\bibfnamefont {M.}~\bibnamefont {Fagotti}},\ }\bibfield
  {title} {\enquote {\bibinfo {title} {{Transport in Out-of-Equilibrium $XXZ$
  Chains: Exact Profiles of Charges and Currents}},}\ }\href {\doibase
  10.1103/PhysRevLett.117.207201} {\bibfield  {journal} {\bibinfo  {journal}
  {Phys. Rev. Lett.}\ }\textbf {\bibinfo {volume} {117}},\ \bibinfo {pages}
  {207201} (\bibinfo {year} {2016})}\BibitemShut {NoStop}%
\bibitem [{\citenamefont {Ran}\ \emph {et~al.}(2020)\citenamefont {Ran},
  \citenamefont {Tirrito}, \citenamefont {Peng}, \citenamefont {Chen},
  \citenamefont {Tagliacozzo}, \citenamefont {Su},\ and\ \citenamefont
  {Lewenstein}}]{Ran2020Tensor}%
  \BibitemOpen
  \bibfield  {author} {\bibinfo {author} {\bibfnamefont {S.-J.}\ \bibnamefont
  {Ran}}, \bibinfo {author} {\bibfnamefont {E.}~\bibnamefont {Tirrito}},
  \bibinfo {author} {\bibfnamefont {C.}~\bibnamefont {Peng}}, \bibinfo {author}
  {\bibfnamefont {X.}~\bibnamefont {Chen}}, \bibinfo {author} {\bibfnamefont
  {L.}~\bibnamefont {Tagliacozzo}}, \bibinfo {author} {\bibfnamefont
  {G.}~\bibnamefont {Su}}, \ and\ \bibinfo {author} {\bibfnamefont
  {M.}~\bibnamefont {Lewenstein}},\ }\href@noop {} {\emph {\bibinfo {title}
  {{Tensor Network Contractions Methods and Applications to Quantum Many-Body
  Systems}}}},\ \bibinfo {edition} {1st}\ ed.,\ Lecture Notes in Physics, 964\
  (\bibinfo  {publisher} {Springer Nature},\ \bibinfo {address} {Cham},\
  \bibinfo {year} {2020})\BibitemShut {NoStop}%
\bibitem [{\citenamefont {Moeckel}\ and\ \citenamefont
  {Kehrein}(2008)}]{Moeckel2008Interaction}%
  \BibitemOpen
  \bibfield  {author} {\bibinfo {author} {\bibfnamefont {M.}~\bibnamefont
  {Moeckel}}\ and\ \bibinfo {author} {\bibfnamefont {S.}~\bibnamefont
  {Kehrein}},\ }\bibfield  {title} {\enquote {\bibinfo {title} {{Interaction
  Quench in the Hubbard Model}},}\ }\href {\doibase
  10.1103/PhysRevLett.100.175702} {\bibfield  {journal} {\bibinfo  {journal}
  {Phys. Rev. Lett.}\ }\textbf {\bibinfo {volume} {100}},\ \bibinfo {pages}
  {175702} (\bibinfo {year} {2008})}\BibitemShut {NoStop}%
\bibitem [{\citenamefont {Rigol}(2009)}]{Rigol2009Breakdown}%
  \BibitemOpen
  \bibfield  {author} {\bibinfo {author} {\bibfnamefont {M.}~\bibnamefont
  {Rigol}},\ }\bibfield  {title} {\enquote {\bibinfo {title} {{Breakdown of
  Thermalization in Finite One-Dimensional Systems}},}\ }\href {\doibase
  10.1103/PhysRevLett.103.100403} {\bibfield  {journal} {\bibinfo  {journal}
  {Phys. Rev. Lett.}\ }\textbf {\bibinfo {volume} {103}},\ \bibinfo {pages}
  {100403} (\bibinfo {year} {2009})}\BibitemShut {NoStop}%
\bibitem [{\citenamefont {Bertini}\ \emph {et~al.}(2015)\citenamefont
  {Bertini}, \citenamefont {Essler}, \citenamefont {Groha},\ and\ \citenamefont
  {Robinson}}]{Bertini2015Prethermalization}%
  \BibitemOpen
  \bibfield  {author} {\bibinfo {author} {\bibfnamefont {B.}~\bibnamefont
  {Bertini}}, \bibinfo {author} {\bibfnamefont {F.~H.~L.}\ \bibnamefont
  {Essler}}, \bibinfo {author} {\bibfnamefont {S.}~\bibnamefont {Groha}}, \
  and\ \bibinfo {author} {\bibfnamefont {N.~J.}\ \bibnamefont {Robinson}},\
  }\bibfield  {title} {\enquote {\bibinfo {title} {{Prethermalization and
  Thermalization in Models with Weak Integrability Breaking}},}\ }\href
  {\doibase 10.1103/PhysRevLett.115.180601} {\bibfield  {journal} {\bibinfo
  {journal} {Phys. Rev. Lett.}\ }\textbf {\bibinfo {volume} {115}},\ \bibinfo
  {pages} {180601} (\bibinfo {year} {2015})}\BibitemShut {NoStop}%
\bibitem [{\citenamefont {Babadi}\ \emph {et~al.}(2015)\citenamefont {Babadi},
  \citenamefont {Demler},\ and\ \citenamefont {Knap}}]{Babadi2015Far}%
  \BibitemOpen
  \bibfield  {author} {\bibinfo {author} {\bibfnamefont {M.}~\bibnamefont
  {Babadi}}, \bibinfo {author} {\bibfnamefont {E.}~\bibnamefont {Demler}}, \
  and\ \bibinfo {author} {\bibfnamefont {M.}~\bibnamefont {Knap}},\ }\bibfield
  {title} {\enquote {\bibinfo {title} {{Far-from-Equilibrium Field Theory of
  Many-Body Quantum Spin Systems: Prethermalization and Relaxation of Spin
  Spiral States in Three Dimensions}},}\ }\href {\doibase
  10.1103/PhysRevX.5.041005} {\bibfield  {journal} {\bibinfo  {journal} {Phys.
  Rev. X}\ }\textbf {\bibinfo {volume} {5}},\ \bibinfo {pages} {041005}
  (\bibinfo {year} {2015})}\BibitemShut {NoStop}%
\bibitem [{\citenamefont {Langen}\ \emph {et~al.}(2016)\citenamefont {Langen},
  \citenamefont {Gasenzer},\ and\ \citenamefont
  {Schmiedmayer}}]{Langen2016Prethermalization}%
  \BibitemOpen
  \bibfield  {author} {\bibinfo {author} {\bibfnamefont {T.}~\bibnamefont
  {Langen}}, \bibinfo {author} {\bibfnamefont {T.}~\bibnamefont {Gasenzer}}, \
  and\ \bibinfo {author} {\bibfnamefont {J.}~\bibnamefont {Schmiedmayer}},\
  }\bibfield  {title} {\enquote {\bibinfo {title} {{Prethermalization and
  universal dynamics in near-integrable quantum systems}},}\ }\href {\doibase
  10.1088/1742-5468/2016/06/064009} {\bibfield  {journal} {\bibinfo  {journal}
  {Journal of Statistical Mechanics: Theory and Experiment}\ }\textbf {\bibinfo
  {volume} {2016}},\ \bibinfo {pages} {064009} (\bibinfo {year}
  {2016})}\BibitemShut {NoStop}%
\bibitem [{\citenamefont {Abanin}\ \emph {et~al.}(2017)\citenamefont {Abanin},
  \citenamefont {De~Roeck}, \citenamefont {Ho},\ and\ \citenamefont
  {Huveneers}}]{Abanin2017A}%
  \BibitemOpen
  \bibfield  {author} {\bibinfo {author} {\bibfnamefont {D.}~\bibnamefont
  {Abanin}}, \bibinfo {author} {\bibfnamefont {W.}~\bibnamefont {De~Roeck}},
  \bibinfo {author} {\bibfnamefont {W.~W.}\ \bibnamefont {Ho}}, \ and\ \bibinfo
  {author} {\bibfnamefont {F.}~\bibnamefont {Huveneers}},\ }\bibfield  {title}
  {\enquote {\bibinfo {title} {{A Rigorous Theory of Many-Body
  Prethermalization for Periodically Driven and Closed Quantum Systems}},}\
  }\href {\doibase 10.1007/s00220-017-2930-x} {\bibfield  {journal} {\bibinfo
  {journal} {Communications in Mathematical Physics}\ }\textbf {\bibinfo
  {volume} {354}},\ \bibinfo {pages} {809} (\bibinfo {year}
  {2017})}\BibitemShut {NoStop}%
\bibitem [{\citenamefont {Alba}\ and\ \citenamefont
  {Fagotti}(2017)}]{Alba2017Prethermalization}%
  \BibitemOpen
  \bibfield  {author} {\bibinfo {author} {\bibfnamefont {V.}~\bibnamefont
  {Alba}}\ and\ \bibinfo {author} {\bibfnamefont {M.}~\bibnamefont {Fagotti}},\
  }\bibfield  {title} {\enquote {\bibinfo {title} {{Prethermalization at Low
  Temperature: The Scent of Long-Range Order}},}\ }\href {\doibase
  10.1103/PhysRevLett.119.010601} {\bibfield  {journal} {\bibinfo  {journal}
  {Phys. Rev. Lett.}\ }\textbf {\bibinfo {volume} {119}},\ \bibinfo {pages}
  {010601} (\bibinfo {year} {2017})}\BibitemShut {NoStop}%
\bibitem [{\citenamefont {Reimann}\ and\ \citenamefont
  {Dabelow}(2019)}]{Reimann2019Typicality}%
  \BibitemOpen
  \bibfield  {author} {\bibinfo {author} {\bibfnamefont {P.}~\bibnamefont
  {Reimann}}\ and\ \bibinfo {author} {\bibfnamefont {L.}~\bibnamefont
  {Dabelow}},\ }\bibfield  {title} {\enquote {\bibinfo {title} {{Typicality of
  Prethermalization}},}\ }\href {\doibase 10.1103/PhysRevLett.122.080603}
  {\bibfield  {journal} {\bibinfo  {journal} {Phys. Rev. Lett.}\ }\textbf
  {\bibinfo {volume} {122}},\ \bibinfo {pages} {080603} (\bibinfo {year}
  {2019})}\BibitemShut {NoStop}%
\bibitem [{\citenamefont {Durnin}\ \emph {et~al.}(2021)\citenamefont {Durnin},
  \citenamefont {Bhaseen},\ and\ \citenamefont
  {Doyon}}]{Durnin2021Nonequilibrium}%
  \BibitemOpen
  \bibfield  {author} {\bibinfo {author} {\bibfnamefont {J.}~\bibnamefont
  {Durnin}}, \bibinfo {author} {\bibfnamefont {M.~J.}\ \bibnamefont {Bhaseen}},
  \ and\ \bibinfo {author} {\bibfnamefont {B.}~\bibnamefont {Doyon}},\
  }\bibfield  {title} {\enquote {\bibinfo {title} {{Nonequilibrium Dynamics and
  Weakly Broken Integrability}},}\ }\href {\doibase
  10.1103/PhysRevLett.127.130601} {\bibfield  {journal} {\bibinfo  {journal}
  {Phys. Rev. Lett.}\ }\textbf {\bibinfo {volume} {127}},\ \bibinfo {pages}
  {130601} (\bibinfo {year} {2021})}\BibitemShut {NoStop}%
\bibitem [{\citenamefont {Fagotti}(2014)}]{Fagotti2014On}%
  \BibitemOpen
  \bibfield  {author} {\bibinfo {author} {\bibfnamefont {M.}~\bibnamefont
  {Fagotti}},\ }\bibfield  {title} {\enquote {\bibinfo {title} {{On
  conservation laws, relaxation and pre-relaxation after a quantum quench}},}\
  }\href {\doibase 10.1088/1742-5468/2014/03/P03016} {\bibfield  {journal}
  {\bibinfo  {journal} {Journal of Statistical Mechanics: Theory and
  Experiment}\ }\textbf {\bibinfo {volume} {2014}},\ \bibinfo {pages} {P03016}
  (\bibinfo {year} {2014})}\BibitemShut {NoStop}%
\bibitem [{\citenamefont {Bertini}\ and\ \citenamefont
  {Fagotti}(2015)}]{Bertini2015pre-relaxation}%
  \BibitemOpen
  \bibfield  {author} {\bibinfo {author} {\bibfnamefont {B.}~\bibnamefont
  {Bertini}}\ and\ \bibinfo {author} {\bibfnamefont {M.}~\bibnamefont
  {Fagotti}},\ }\bibfield  {title} {\enquote {\bibinfo {title} {{Pre-relaxation
  in weakly interacting models}},}\ }\href {\doibase
  10.1088/1742-5468/2015/07/P07012} {\bibfield  {journal} {\bibinfo  {journal}
  {Journal of Statistical Mechanics: Theory and Experiment}\ }\textbf {\bibinfo
  {volume} {2015}},\ \bibinfo {pages} {P07012} (\bibinfo {year}
  {2015})}\BibitemShut {NoStop}%
\bibitem [{\citenamefont {Zadnik}\ and\ \citenamefont
  {Garrahan}(2023)}]{zadnik2023slow}%
  \BibitemOpen
  \bibfield  {author} {\bibinfo {author} {\bibfnamefont {L.}~\bibnamefont
  {Zadnik}}\ and\ \bibinfo {author} {\bibfnamefont {J.~P.}\ \bibnamefont
  {Garrahan}},\ }\href {\doibase 10.48550/arXiv.2304.10394} {\enquote {\bibinfo
  {title} {{Slow heterogeneous relaxation due to constraints in dual XXZ
  models}},}\ } (\bibinfo {year} {2023}),\ \Eprint
  {http://arxiv.org/abs/2304.10394}{arXiv:2304.10394
  [cond-mat.stat-mech]}\BibitemShut {NoStop}%
\bibitem [{\citenamefont {MacDonald}\ \emph {et~al.}(1988)\citenamefont
  {MacDonald}, \citenamefont {Girvin},\ and\ \citenamefont
  {Yoshioka}}]{MacDonald1988tU}%
  \BibitemOpen
  \bibfield  {author} {\bibinfo {author} {\bibfnamefont {A.~H.}\ \bibnamefont
  {MacDonald}}, \bibinfo {author} {\bibfnamefont {S.~M.}\ \bibnamefont
  {Girvin}}, \ and\ \bibinfo {author} {\bibfnamefont {D.}~\bibnamefont
  {Yoshioka}},\ }\bibfield  {title} {\enquote {\bibinfo {title} {{$\frac{t}{U}$
  expansion for the Hubbard model}},}\ }\href {\doibase
  10.1103/PhysRevB.37.9753} {\bibfield  {journal} {\bibinfo  {journal} {Phys.
  Rev. B}\ }\textbf {\bibinfo {volume} {37}},\ \bibinfo {pages} {9753}
  (\bibinfo {year} {1988})}\BibitemShut {NoStop}%
\bibitem [{\citenamefont {Zadnik}\ and\ \citenamefont
  {Fagotti}(2021)}]{Zadnik2021The}%
  \BibitemOpen
  \bibfield  {author} {\bibinfo {author} {\bibfnamefont {L.}~\bibnamefont
  {Zadnik}}\ and\ \bibinfo {author} {\bibfnamefont {M.}~\bibnamefont
  {Fagotti}},\ }\bibfield  {title} {\enquote {\bibinfo {title} {{The Folded
  Spin-1/2 XXZ Model: I. Diagonalisation, Jamming, and Ground State
  Properties}},}\ }\href {\doibase 10.21468/SciPostPhysCore.4.2.010} {\bibfield
   {journal} {\bibinfo  {journal} {SciPost Phys. Core}\ }\textbf {\bibinfo
  {volume} {4}},\ \bibinfo {pages} {010} (\bibinfo {year} {2021})}\BibitemShut
  {NoStop}%
\bibitem [{\citenamefont {Moudgalya}\ and\ \citenamefont
  {Motrunich}(2022)}]{Moudgalya2022Hilbert}%
  \BibitemOpen
  \bibfield  {author} {\bibinfo {author} {\bibfnamefont {S.}~\bibnamefont
  {Moudgalya}}\ and\ \bibinfo {author} {\bibfnamefont {O.~I.}\ \bibnamefont
  {Motrunich}},\ }\bibfield  {title} {\enquote {\bibinfo {title} {{Hilbert
  Space Fragmentation and Commutant Algebras}},}\ }\href {\doibase
  10.1103/PhysRevX.12.011050} {\bibfield  {journal} {\bibinfo  {journal} {Phys.
  Rev. X}\ }\textbf {\bibinfo {volume} {12}},\ \bibinfo {pages} {011050}
  (\bibinfo {year} {2022})}\BibitemShut {NoStop}%
\bibitem [{\citenamefont {Biroli}\ \emph {et~al.}(2008)\citenamefont {Biroli},
  \citenamefont {Chamon},\ and\ \citenamefont {Zamponi}}]{Biroli2008Theory}%
  \BibitemOpen
  \bibfield  {author} {\bibinfo {author} {\bibfnamefont {G.}~\bibnamefont
  {Biroli}}, \bibinfo {author} {\bibfnamefont {C.}~\bibnamefont {Chamon}}, \
  and\ \bibinfo {author} {\bibfnamefont {F.}~\bibnamefont {Zamponi}},\
  }\bibfield  {title} {\enquote {\bibinfo {title} {{Theory of the superglass
  phase}},}\ }\href {\doibase 10.1103/PhysRevB.78.224306} {\bibfield  {journal}
  {\bibinfo  {journal} {Phys. Rev. B}\ }\textbf {\bibinfo {volume} {78}},\
  \bibinfo {pages} {224306} (\bibinfo {year} {2008})}\BibitemShut {NoStop}%
\bibitem [{\citenamefont {Nussinov}\ \emph {et~al.}(2013)\citenamefont
  {Nussinov}, \citenamefont {Johnson}, \citenamefont {Graf},\ and\
  \citenamefont {Balatsky}}]{Nussinov2013Mapping}%
  \BibitemOpen
  \bibfield  {author} {\bibinfo {author} {\bibfnamefont {Z.}~\bibnamefont
  {Nussinov}}, \bibinfo {author} {\bibfnamefont {P.}~\bibnamefont {Johnson}},
  \bibinfo {author} {\bibfnamefont {M.~J.}\ \bibnamefont {Graf}}, \ and\
  \bibinfo {author} {\bibfnamefont {A.~V.}\ \bibnamefont {Balatsky}},\
  }\bibfield  {title} {\enquote {\bibinfo {title} {{Mapping between finite
  temperature classical and zero temperature quantum systems: Quantum critical
  jamming and quantum dynamical heterogeneities}},}\ }\href {\doibase
  10.1103/PhysRevB.87.184202} {\bibfield  {journal} {\bibinfo  {journal} {Phys.
  Rev. B}\ }\textbf {\bibinfo {volume} {87}},\ \bibinfo {pages} {184202}
  (\bibinfo {year} {2013})}\BibitemShut {NoStop}%
\bibitem [{\citenamefont {Artiaco}\ \emph {et~al.}(2021)\citenamefont
  {Artiaco}, \citenamefont {Balducci}, \citenamefont {Parisi},\ and\
  \citenamefont {Scardicchio}}]{Artiaco2021Quantum}%
  \BibitemOpen
  \bibfield  {author} {\bibinfo {author} {\bibfnamefont {C.}~\bibnamefont
  {Artiaco}}, \bibinfo {author} {\bibfnamefont {F.}~\bibnamefont {Balducci}},
  \bibinfo {author} {\bibfnamefont {G.}~\bibnamefont {Parisi}}, \ and\ \bibinfo
  {author} {\bibfnamefont {A.}~\bibnamefont {Scardicchio}},\ }\bibfield
  {title} {\enquote {\bibinfo {title} {{Quantum jamming: Critical properties of
  a quantum mechanical perceptron}},}\ }\href {\doibase
  10.1103/PhysRevA.103.L040203} {\bibfield  {journal} {\bibinfo  {journal}
  {Phys. Rev. A}\ }\textbf {\bibinfo {volume} {103}},\ \bibinfo {pages}
  {L040203} (\bibinfo {year} {2021})}\BibitemShut {NoStop}%
\bibitem [{\citenamefont {Biroli}(2007)}]{Biroli2007A}%
  \BibitemOpen
  \bibfield  {author} {\bibinfo {author} {\bibfnamefont {G.}~\bibnamefont
  {Biroli}},\ }\bibfield  {title} {\enquote {\bibinfo {title} {{A new kind of
  phase transition?}}}\ }\href {\doibase 10.1038/nphys580} {\bibfield
  {journal} {\bibinfo  {journal} {Nature Physics}\ }\textbf {\bibinfo {volume}
  {3}},\ \bibinfo {pages} {222} (\bibinfo {year} {2007})}\BibitemShut {NoStop}%
\bibitem [{\citenamefont {Weitenberg}\ \emph {et~al.}(2011)\citenamefont
  {Weitenberg}, \citenamefont {Endres}, \citenamefont {Sherson}, \citenamefont
  {Cheneau}, \citenamefont {Schau{\ss}}, \citenamefont {Fukuhara},
  \citenamefont {Bloch},\ and\ \citenamefont {Kuhr}}]{Weitenberg2011Single}%
  \BibitemOpen
  \bibfield  {author} {\bibinfo {author} {\bibfnamefont {C.}~\bibnamefont
  {Weitenberg}}, \bibinfo {author} {\bibfnamefont {M.}~\bibnamefont {Endres}},
  \bibinfo {author} {\bibfnamefont {J.~F.}\ \bibnamefont {Sherson}}, \bibinfo
  {author} {\bibfnamefont {M.}~\bibnamefont {Cheneau}}, \bibinfo {author}
  {\bibfnamefont {P.}~\bibnamefont {Schau{\ss}}}, \bibinfo {author}
  {\bibfnamefont {T.}~\bibnamefont {Fukuhara}}, \bibinfo {author}
  {\bibfnamefont {I.}~\bibnamefont {Bloch}}, \ and\ \bibinfo {author}
  {\bibfnamefont {S.}~\bibnamefont {Kuhr}},\ }\bibfield  {title} {\enquote
  {\bibinfo {title} {{Single-spin addressing in an atomic Mott insulator}},}\
  }\href {\doibase 10.1038/nature09827} {\bibfield  {journal} {\bibinfo
  {journal} {Nature}\ }\textbf {\bibinfo {volume} {471}},\ \bibinfo {pages}
  {319} (\bibinfo {year} {2011})}\BibitemShut {NoStop}%
\bibitem [{\citenamefont {Došlić}\ \emph {et~al.}(2023)\citenamefont
  {Došlić}, \citenamefont {Puljiz}, \citenamefont {Šebek},\ and\
  \citenamefont {Žubrinić}}]{doslic2023complexity}%
  \BibitemOpen
  \bibfield  {author} {\bibinfo {author} {\bibfnamefont {T.}~\bibnamefont
  {Došlić}}, \bibinfo {author} {\bibfnamefont {M.}~\bibnamefont {Puljiz}},
  \bibinfo {author} {\bibfnamefont {S.}~\bibnamefont {Šebek}}, \ and\ \bibinfo
  {author} {\bibfnamefont {J.}~\bibnamefont {Žubrinić}},\ }\href {\doibase
  10.48550/arXiv.2302.08791} {\enquote {\bibinfo {title} {{Complexity Function
  of Jammed Configurations of Rydberg Atoms}},}\ } (\bibinfo {year} {2023}),\
  \Eprint {http://arxiv.org/abs/2302.08791}{arXiv:2302.08791
  [math.CO]}\BibitemShut {NoStop}%
\bibitem [{\citenamefont {Gotta}\ \emph {et~al.}(2021)\citenamefont {Gotta},
  \citenamefont {Mazza}, \citenamefont {Simon},\ and\ \citenamefont
  {Roux}}]{Gotta2021Two}%
  \BibitemOpen
  \bibfield  {author} {\bibinfo {author} {\bibfnamefont {L.}~\bibnamefont
  {Gotta}}, \bibinfo {author} {\bibfnamefont {L.}~\bibnamefont {Mazza}},
  \bibinfo {author} {\bibfnamefont {P.}~\bibnamefont {Simon}}, \ and\ \bibinfo
  {author} {\bibfnamefont {G.}~\bibnamefont {Roux}},\ }\bibfield  {title}
  {\enquote {\bibinfo {title} {{Two-Fluid Coexistence in a Spinless Fermions
  Chain with Pair Hopping}},}\ }\href {\doibase 10.1103/PhysRevLett.126.206805}
  {\bibfield  {journal} {\bibinfo  {journal} {Phys. Rev. Lett.}\ }\textbf
  {\bibinfo {volume} {126}},\ \bibinfo {pages} {206805} (\bibinfo {year}
  {2021})}\BibitemShut {NoStop}%
\bibitem [{\citenamefont {Krajnik}\ \emph {et~al.}(2023)\citenamefont
  {Krajnik}, \citenamefont {Schmidt}, \citenamefont {Pasquier}, \citenamefont
  {Prosen},\ and\ \citenamefont {Ilievski}}]{krajnik2023universal}%
  \BibitemOpen
  \bibfield  {author} {\bibinfo {author} {\bibfnamefont {{\v{Z}}.}~\bibnamefont
  {Krajnik}}, \bibinfo {author} {\bibfnamefont {J.}~\bibnamefont {Schmidt}},
  \bibinfo {author} {\bibfnamefont {V.}~\bibnamefont {Pasquier}}, \bibinfo
  {author} {\bibfnamefont {T.}~\bibnamefont {Prosen}}, \ and\ \bibinfo {author}
  {\bibfnamefont {E.}~\bibnamefont {Ilievski}},\ }\href {\doibase
  10.48550/arXiv.2208.01463} {\enquote {\bibinfo {title} {{Universal anomalous
  fluctuations in charged single-file systems}},}\ } (\bibinfo {year} {2023}),\
  \Eprint {http://arxiv.org/abs/2208.01463}{arXiv:2208.01463
  [cond-mat.stat-mech]}\BibitemShut {NoStop}%
\bibitem [{\citenamefont {Borsi}\ \emph {et~al.}(2023)\citenamefont {Borsi},
  \citenamefont {Pristy\'ak},\ and\ \citenamefont {Pozsgay}}]{Borsi2023Matrix}%
  \BibitemOpen
  \bibfield  {author} {\bibinfo {author} {\bibfnamefont {M.}~\bibnamefont
  {Borsi}}, \bibinfo {author} {\bibfnamefont {L.}~\bibnamefont {Pristy\'ak}}, \
  and\ \bibinfo {author} {\bibfnamefont {B.}~\bibnamefont {Pozsgay}},\
  }\bibfield  {title} {\enquote {\bibinfo {title} {Matrix product symmetries
  and breakdown of thermalization from hard rod deformations},}\ }\href
  {\doibase 10.1103/PhysRevLett.131.037101} {\bibfield  {journal} {\bibinfo
  {journal} {Phys. Rev. Lett.}\ }\textbf {\bibinfo {volume} {131}},\ \bibinfo
  {pages} {037101} (\bibinfo {year} {2023})}\BibitemShut {NoStop}%
\bibitem [{\citenamefont {Pozsgay}\ \emph {et~al.}(2021)\citenamefont
  {Pozsgay}, \citenamefont {Gombor},\ and\ \citenamefont
  {Hutsalyuk}}]{Pozsgay2021Integrable}%
  \BibitemOpen
  \bibfield  {author} {\bibinfo {author} {\bibfnamefont {B.}~\bibnamefont
  {Pozsgay}}, \bibinfo {author} {\bibfnamefont {T.}~\bibnamefont {Gombor}}, \
  and\ \bibinfo {author} {\bibfnamefont {A.}~\bibnamefont {Hutsalyuk}},\
  }\bibfield  {title} {\enquote {\bibinfo {title} {{Integrable hard-rod
  deformation of the Heisenberg spin chains}},}\ }\href {\doibase
  10.1103/PhysRevE.104.064124} {\bibfield  {journal} {\bibinfo  {journal}
  {Phys. Rev. E}\ }\textbf {\bibinfo {volume} {104}},\ \bibinfo {pages}
  {064124} (\bibinfo {year} {2021})}\BibitemShut {NoStop}%
\bibitem [{\citenamefont {Bariev}(1991)}]{Bariev1991Integrable}%
  \BibitemOpen
  \bibfield  {author} {\bibinfo {author} {\bibfnamefont {R.~Z.}\ \bibnamefont
  {Bariev}},\ }\bibfield  {title} {\enquote {\bibinfo {title} {{Integrable spin
  chain with two- and three-particle interactions}},}\ }\href {\doibase
  10.1088/0305-4470/24/10/010} {\bibfield  {journal} {\bibinfo  {journal}
  {Journal of Physics A: Mathematical and General}\ }\textbf {\bibinfo {volume}
  {24}},\ \bibinfo {pages} {L549} (\bibinfo {year} {1991})}\BibitemShut
  {NoStop}%
\bibitem [{\citenamefont {Bidzhiev}\ \emph {et~al.}(2022)\citenamefont
  {Bidzhiev}, \citenamefont {Fagotti},\ and\ \citenamefont
  {Zadnik}}]{Bidzhiev2022Macroscopic}%
  \BibitemOpen
  \bibfield  {author} {\bibinfo {author} {\bibfnamefont {K.}~\bibnamefont
  {Bidzhiev}}, \bibinfo {author} {\bibfnamefont {M.}~\bibnamefont {Fagotti}}, \
  and\ \bibinfo {author} {\bibfnamefont {L.}~\bibnamefont {Zadnik}},\
  }\bibfield  {title} {\enquote {\bibinfo {title} {{Macroscopic Effects of
  Localized Measurements in Jammed States of Quantum Spin Chains}},}\ }\href
  {\doibase 10.1103/PhysRevLett.128.130603} {\bibfield  {journal} {\bibinfo
  {journal} {Phys. Rev. Lett.}\ }\textbf {\bibinfo {volume} {128}},\ \bibinfo
  {pages} {130603} (\bibinfo {year} {2022})}\BibitemShut {NoStop}%
\bibitem [{\citenamefont {Zadnik}\ \emph {et~al.}(2022)\citenamefont {Zadnik},
  \citenamefont {Bocini}, \citenamefont {Bidzhiev},\ and\ \citenamefont
  {Fagotti}}]{Zadnik2022Measurement}%
  \BibitemOpen
  \bibfield  {author} {\bibinfo {author} {\bibfnamefont {L.}~\bibnamefont
  {Zadnik}}, \bibinfo {author} {\bibfnamefont {S.}~\bibnamefont {Bocini}},
  \bibinfo {author} {\bibfnamefont {K.}~\bibnamefont {Bidzhiev}}, \ and\
  \bibinfo {author} {\bibfnamefont {M.}~\bibnamefont {Fagotti}},\ }\bibfield
  {title} {\enquote {\bibinfo {title} {{Measurement catastrophe and ballistic
  spread of charge density with vanishing current}},}\ }\href {\doibase
  10.1088/1751-8121/aca254} {\bibfield  {journal} {\bibinfo  {journal} {Journal
  of Physics A: Mathematical and Theoretical}\ }\textbf {\bibinfo {volume}
  {55}},\ \bibinfo {pages} {474001} (\bibinfo {year} {2022})}\BibitemShut
  {NoStop}%
\bibitem [{\citenamefont {Bethe}(1931)}]{Bethe1931Zur}%
  \BibitemOpen
  \bibfield  {author} {\bibinfo {author} {\bibfnamefont {H.}~\bibnamefont
  {Bethe}},\ }\bibfield  {title} {\enquote {\bibinfo {title} {{Zur Theorie der
  Metalle}},}\ }\href {\doibase 10.1007/BF01341708} {\bibfield  {journal}
  {\bibinfo  {journal} {Zeitschrift f{\"u}r Physik}\ }\textbf {\bibinfo
  {volume} {71}},\ \bibinfo {pages} {205} (\bibinfo {year} {1931})}\BibitemShut
  {NoStop}%
\bibitem [{\citenamefont {Bonnes}\ \emph {et~al.}(2014)\citenamefont {Bonnes},
  \citenamefont {Essler},\ and\ \citenamefont {L\"auchli}}]{Bonnes2014Light}%
  \BibitemOpen
  \bibfield  {author} {\bibinfo {author} {\bibfnamefont {L.}~\bibnamefont
  {Bonnes}}, \bibinfo {author} {\bibfnamefont {F.~H.~L.}\ \bibnamefont
  {Essler}}, \ and\ \bibinfo {author} {\bibfnamefont {A.~M.}\ \bibnamefont
  {L\"auchli}},\ }\bibfield  {title} {\enquote {\bibinfo {title}
  {{``Light-Cone'' Dynamics After Quantum Quenches in Spin Chains}},}\ }\href
  {\doibase 10.1103/PhysRevLett.113.187203} {\bibfield  {journal} {\bibinfo
  {journal} {Phys. Rev. Lett.}\ }\textbf {\bibinfo {volume} {113}},\ \bibinfo
  {pages} {187203} (\bibinfo {year} {2014})}\BibitemShut {NoStop}%
\bibitem [{\citenamefont {Anderson}(1958)}]{Anderson1958Absence}%
  \BibitemOpen
  \bibfield  {author} {\bibinfo {author} {\bibfnamefont {P.~W.}\ \bibnamefont
  {Anderson}},\ }\bibfield  {title} {\enquote {\bibinfo {title} {{Absence of
  Diffusion in Certain Random Lattices}},}\ }\href {\doibase
  10.1103/PhysRev.109.1492} {\bibfield  {journal} {\bibinfo  {journal} {Phys.
  Rev.}\ }\textbf {\bibinfo {volume} {109}},\ \bibinfo {pages} {1492} (\bibinfo
  {year} {1958})}\BibitemShut {NoStop}%
\bibitem [{\citenamefont {Bocini}\ and\ \citenamefont
  {Fagotti}(2023)}]{bocini2023growing}%
  \BibitemOpen
  \bibfield  {author} {\bibinfo {author} {\bibfnamefont {S.}~\bibnamefont
  {Bocini}}\ and\ \bibinfo {author} {\bibfnamefont {M.}~\bibnamefont
  {Fagotti}},\ }\href {\doibase 10.48550/arXiv.2210.15585} {\enquote {\bibinfo
  {title} {{Growing Schr\"odinger's cat states by local unitary time evolution
  of product states}},}\ } (\bibinfo {year} {2023}),\ \Eprint
  {http://arxiv.org/abs/2210.15585}{arXiv:2210.15585 [quant-ph]}\BibitemShut
  {NoStop}%
\bibitem [{\citenamefont {Lieb}\ and\ \citenamefont
  {Robinson}(1972)}]{Lieb1972The}%
  \BibitemOpen
  \bibfield  {author} {\bibinfo {author} {\bibfnamefont {E.~H.}\ \bibnamefont
  {Lieb}}\ and\ \bibinfo {author} {\bibfnamefont {D.~W.}\ \bibnamefont
  {Robinson}},\ }\bibfield  {title} {\enquote {\bibinfo {title} {{The finite
  group velocity of quantum spin systems}},}\ }\href@noop {} {\bibfield
  {journal} {\bibinfo  {journal} {Communications in Mathematical Physics}\
  }\textbf {\bibinfo {volume} {28}},\ \bibinfo {pages} {251 } (\bibinfo {year}
  {1972})}\BibitemShut {NoStop}%
\bibitem [{\citenamefont {Fröwis}\ and\ \citenamefont
  {Dür}(2012)}]{Frowis2012Measures}%
  \BibitemOpen
  \bibfield  {author} {\bibinfo {author} {\bibfnamefont {F.}~\bibnamefont
  {Fröwis}}\ and\ \bibinfo {author} {\bibfnamefont {W.}~\bibnamefont {Dür}},\
  }\bibfield  {title} {\enquote {\bibinfo {title} {{Measures of macroscopicity
  for quantum spin systems}},}\ }\href {\doibase 10.1088/1367-2630/14/9/093039}
  {\bibfield  {journal} {\bibinfo  {journal} {New Journal of Physics}\ }\textbf
  {\bibinfo {volume} {14}},\ \bibinfo {pages} {093039} (\bibinfo {year}
  {2012})}\BibitemShut {NoStop}%
\bibitem [{\citenamefont {Zauner}\ \emph {et~al.}(2015)\citenamefont {Zauner},
  \citenamefont {Ganahl}, \citenamefont {Evertz},\ and\ \citenamefont
  {Nishino}}]{Zauner2015Time}%
  \BibitemOpen
  \bibfield  {author} {\bibinfo {author} {\bibfnamefont {V.}~\bibnamefont
  {Zauner}}, \bibinfo {author} {\bibfnamefont {M.}~\bibnamefont {Ganahl}},
  \bibinfo {author} {\bibfnamefont {H.~G.}\ \bibnamefont {Evertz}}, \ and\
  \bibinfo {author} {\bibfnamefont {T.}~\bibnamefont {Nishino}},\ }\bibfield
  {title} {\enquote {\bibinfo {title} {{Time evolution within a comoving
  window: scaling of signal fronts and magnetization plateaus after a local
  quench in quantum spin chains}},}\ }\href {\doibase
  10.1088/0953-8984/27/42/425602} {\bibfield  {journal} {\bibinfo  {journal}
  {Journal of Physics: Condensed Matter}\ }\textbf {\bibinfo {volume} {27}},\
  \bibinfo {pages} {425602} (\bibinfo {year} {2015})}\BibitemShut {NoStop}%
\bibitem [{\citenamefont {Eisler}\ and\ \citenamefont
  {Maislinger}(2020)}]{Eisler2020Front}%
  \BibitemOpen
  \bibfield  {author} {\bibinfo {author} {\bibfnamefont {V.}~\bibnamefont
  {Eisler}}\ and\ \bibinfo {author} {\bibfnamefont {F.}~\bibnamefont
  {Maislinger}},\ }\bibfield  {title} {\enquote {\bibinfo {title} {{Front
  dynamics in the XY chain after local excitations}},}\ }\href {\doibase
  10.21468/SciPostPhys.8.3.037} {\bibfield  {journal} {\bibinfo  {journal}
  {SciPost Phys.}\ }\textbf {\bibinfo {volume} {8}},\ \bibinfo {pages} {037}
  (\bibinfo {year} {2020})}\BibitemShut {NoStop}%
\bibitem [{\citenamefont {Fagotti}(2022)}]{Fagotti2022Global}%
  \BibitemOpen
  \bibfield  {author} {\bibinfo {author} {\bibfnamefont {M.}~\bibnamefont
  {Fagotti}},\ }\bibfield  {title} {\enquote {\bibinfo {title} {{Global
  Quenches after Localized Perturbations}},}\ }\href {\doibase
  10.1103/PhysRevLett.128.110602} {\bibfield  {journal} {\bibinfo  {journal}
  {Phys. Rev. Lett.}\ }\textbf {\bibinfo {volume} {128}},\ \bibinfo {pages}
  {110602} (\bibinfo {year} {2022})}\BibitemShut {NoStop}%
\bibitem [{\citenamefont {Fagotti}\ \emph {et~al.}(2022)\citenamefont
  {Fagotti}, \citenamefont {Marić},\ and\ \citenamefont
  {Zadnik}}]{fagotti2022nonequilibrium}%
  \BibitemOpen
  \bibfield  {author} {\bibinfo {author} {\bibfnamefont {M.}~\bibnamefont
  {Fagotti}}, \bibinfo {author} {\bibfnamefont {V.}~\bibnamefont {Marić}}, \
  and\ \bibinfo {author} {\bibfnamefont {L.}~\bibnamefont {Zadnik}},\ }\href
  {\doibase 10.48550/arXiv.2205.02221} {\enquote {\bibinfo {title}
  {{Nonequilibrium symmetry-protected topological order: emergence of semilocal
  Gibbs ensembles}},}\ } (\bibinfo {year} {2022}),\ \Eprint
  {http://arxiv.org/abs/2205.02221}{arXiv:2205.02221
  [cond-mat.stat-mech]}\BibitemShut {NoStop}%
\end{thebibliography}%
\end{document}